\documentclass[%
 reprint,
 amsmath,amssymb,
 aps,
]{revtex4-2}

\usepackage[colorlinks=true,allcolors=blue]{hyperref} % blue cross-refs
\usepackage{graphicx}% Include figure files
\usepackage{dcolumn}% Align table columns on decimal point
\usepackage{bm}% bold math
\usepackage[mathlines]{lineno}% Enable numbering of text and display math
\usepackage{physics} % Physics symbols

\begin{document}

\preprint{APS/123-QED}

\title{Finite-temperature electron-capture rates for neutron-rich nuclei around N=50 and effects on core-collapse supernovae simulations}% Force line breaks with \\
%\thanks{A footnote to the article title}%

\author{S. Giraud}
% \email[]{giraud@frib.msu.edu}
\affiliation{National Superconducting Cyclotron Laboratory, Michigan State University, East Lansing, MI 48824, USA\looseness=-1}
\affiliation{Joint Institute for Nuclear Astrophysics: Center for the Evolution of the Elements, Michigan State University, East Lansing, MI 48824, USA\looseness=-1}
\affiliation{Department of Physics and Astronomy, Michigan State University, East Lansing, MI 48824, USA\looseness=-1}

\author{R.G.T. Zegers}
% \email[]{zegers@frib.msu.edu}
\affiliation{National Superconducting Cyclotron Laboratory, Michigan State University, East Lansing, MI 48824, USA\looseness=-1}
\affiliation{Joint Institute for Nuclear Astrophysics: Center for the Evolution of the Elements, Michigan State University, East Lansing, MI 48824, USA\looseness=-1}
\affiliation{Department of Physics and Astronomy, Michigan State University, East Lansing, MI 48824, USA\looseness=-1}

\author{B.A. Brown}
% \email[]{}
\affiliation{National Superconducting Cyclotron Laboratory, Michigan State University, East Lansing, MI 48824, USA\looseness=-1}
\affiliation{Joint Institute for Nuclear Astrophysics: Center for the Evolution of the Elements, Michigan State University, East Lansing, MI 48824, USA\looseness=-1}
\affiliation{Department of Physics and Astronomy, Michigan State University, East Lansing, MI 48824, USA\looseness=-1}

\author{J.-M. Gabler}
% \email[]{}
\affiliation{National Superconducting Cyclotron Laboratory, Michigan State University, East Lansing, MI 48824, USA\looseness=-1}
\affiliation{Joint Institute for Nuclear Astrophysics: Center for the Evolution of the Elements, Michigan State University, East Lansing, MI 48824, USA\looseness=-1}
\affiliation{Department of Physics and Astronomy, Michigan State University, East Lansing, MI 48824, USA\looseness=-1}

\author{J. Lesniak}
% \email[]{}
\affiliation{National Superconducting Cyclotron Laboratory, Michigan State University, East Lansing, MI 48824, USA\looseness=-1}
\affiliation{Joint Institute for Nuclear Astrophysics: Center for the Evolution of the Elements, Michigan State University, East Lansing, MI 48824, USA\looseness=-1}
\affiliation{Department of Physics and Astronomy, Michigan State University, East Lansing, MI 48824, USA\looseness=-1}

\author{J. Rebenstock}
% \email[]{}
\affiliation{National Superconducting Cyclotron Laboratory, Michigan State University, East Lansing, MI 48824, USA\looseness=-1}
\affiliation{Joint Institute for Nuclear Astrophysics: Center for the Evolution of the Elements, Michigan State University, East Lansing, MI 48824, USA\looseness=-1}
\affiliation{Department of Physics and Astronomy, Michigan State University, East Lansing, MI 48824, USA\looseness=-1}

\author{E. M. Ney}
% \email[]{evan.ney@unc.edu}
\affiliation{Department of Physics and Astronomy, CB 3255, University of North
Carolina, Chapel Hill, North Carolina 27599-3255, USA\looseness=-1}

\author{J. Engel}
% \email[]{engelj@unc.edu}
\affiliation{Department of Physics and Astronomy, CB 3255, University of North
Carolina, Chapel Hill, North Carolina 27599-3255, USA\looseness=-1}

\author{A. Ravli\'c}
% \email[]{aravlic@phy.hr}
\affiliation{Department of Physics, Faculty of Science, University of Zagreb, Bijeni\v{c}ka c. 32, 10000 Zagreb, Croatia}

\author{N. Paar}
% \email[]{npaar@phy.hr}
\affiliation{Department of Physics, Faculty of Science, University of Zagreb, Bijeni\v{c}ka c. 32, 10000 Zagreb, Croatia}

\date{\today}% It is always \today, today,
             %  but any date may be explicitly specified

\begin{abstract}
The temperature dependence of stellar electron-capture (EC) rates is investigated, with a focus on nuclei around $N=50$, just above $Z=28$, which play an important role during the collapse phase of core-collapse supernovae (CCSN). Two new microscopic calculations of stellar EC rates are obtained from a relativistic and a non-relativistic finite-temperature quasiparticle random-phase approximation approaches, for a conventional grid of temperatures and densities. In both approaches, EC rates due to Gamow-Teller transitions are included. In the relativistic calculation contributions from first-forbidden transitions are also included, and add strongly to the EC rates. The new EC rates are compared with large-scale shell model calculations for the specific case of $^{86}$Kr, providing insight into the finite-temperature effects on the EC rates. At relevant thermodynamic conditions for core-collapse, the discrepancies between the different calculations of this work are within about one order of magnitude. Numerical simulations of CCSN are performed with the spherically-symmetric GR1D simulation code to quantify the impact of such differences on the dynamics of the collapse. These simulations also include EC rates based on two parametrized approximations. A comparison of the neutrino luminosities and enclosed mass at core bounce shows that differences between simulations with different sets of EC rates are relatively small ($\approx 5\%$), suggesting that the EC rates used as inputs for these simulations have become well constrained.

%\begin{description}
%\item[Usage]
%Secondary publications and information retrieval purposes.
%\item[Structure]
%You may use the \texttt{description} environment to structure your abstract;
%use the optional argument of the \verb+\item+ command to give the category of each item. 
%\end{description}
\end{abstract}

%\keywords{Suggested keywords}%Use showkeys class option if keyword
                              %display desired
\maketitle

%\tableofcontents

\section{Introduction}
Electron-capture (EC) rates play a key role in various astrophysical phenomena, such as the final evolution of intermediate-mass stars~\cite{Zha2019,Doherty2017}, core-collapse supernovae (CCSN)~\cite{Fuller1982,Janke2007,Sullivan2015,Pascal2020}, thermal evolution of the neutron-star crust~\cite{Schatz2014,Chamel2021}, and nucleosynthesis in thermonuclear supernovae~\cite{Bravo2019,Iwamoto1999}. For a recent review work the reader may refer to Ref.~\cite{Langanke2021}. CCSN are particularly impacted by the rate of electron captures prior and during the collapse phase as it defines the electron fraction ($Y_e$), which drives the collapse dynamics and sets the diameter of the core at bounce~\cite{Sullivan2015,Pascal2020}. Indeed, at the onset of the collapse, the combination of a high stellar temperature ($T\sim$10 GK), high density ($\rho\sim$10$^{10}$g.cm$^{-3}$), and low entropy ($s\sim$1kB) leads to a nuclear statistical equilibrium~\cite{Bethe1990} in the core. While the core density increases, the electron captures on nuclei and free protons reduce $Y_e$ and produce electron-type neutrinos, which escape the core freely while carrying away energy and entropy. Consequently, $Y_e$ further decreases and the collapse accelerates. The electron-capture reactions on nuclei dominate because the mass fraction of free nucleons is small compared to that of nuclei~\cite{Langanke2000}. Previous studies~\cite{Sullivan2015,Pascal2020} have shown that the nuclei having the largest impact on the evolution of $Y_e$, and therefore on the production of electron neutrinos, are located along the $N=50$ shell closure near $^{78}$Ni and along $N=82$ near $^{128}$Pd. At $\rho\gtrsim$10$^{-12}$ g.cm$^{-3}$, the electron-neutrino diffusion timescale becomes longer than the dynamical timescale of the collapse, the electron neutrinos become trapped, and a $\beta$-equilibrium establishes~\cite{Shapiro2008,Bethe1990}. The core continues its collapse up to $\rho\gtrsim$ n$_{sat}$ $\approx$ $2.81 \cdot 10^{14}$ g.cm$^{-3}$. At the interface where the in-fall velocity is equal to the speed of sound in the medium a shock wave forms and propagates outwards. The mass of the inner core, approximately the Chandrasekhar mass, is proportional to ${Y_e}^2$~\cite{Shapiro2008,Bethe1990}.

During the collapse, the nuclei are in thermal equilibrium and undergo continuum EC. As the stellar density is high, the Fermi energy is also high, and ECs can occur to states in the daughter at relatively high excitation energy. In addition, because the temperature is also high, excited states in the parent are populated and ECs can occur on these states~\cite{Bethe1979}. The EC rates are mediated by Gamow-Teller transitions and forbidden transitions ~\cite{Dzhioev2020,Litvinova2021,Ravlic2020}. The stellar conditions cannot be reproduced in the laboratory and to estimate the rates at extreme thermodynamic conditions one has to rely on theoretical models. The theoretical models must be benchmarked with experimental data where available, i.e. primarily from the ground state of the parent nucleus. While EC/$\beta$+-decay data provide benchmarks, the accessible Q-value window is very limited, especially on the neutron-rich side of stability, which contains the nuclei of most interest in the collapse phase of supernovae. Therefore, GT strengths extracted from (n,p)-type charge-exchange experiments~\cite{Langanke2021} at intermediate energies, such as (n,p)~\cite{Alford1993,Vetterli1989,El-Kateb1994,Williams1995}, (d,$^2$He)~\cite{Ohnuma1993,Xu1995,Frekers2018}, and (t,$^3$He)~\cite{Noji2014,Gao2020,Zamora2019} reactions, have become the most important tool for testing theoretical models. 

Fuller, Fowler, and Newman \cite{FFN1980} (FFN) were the first to perform calculations for a wide grid of stellar conditions and for an ensemble of nuclei near stability with mass number 21$<A<$60. The first FFN formulation was based on strict assumptions where a single resonance contains the total GT strength. The energy of this resonance was determined phenomenologically and the total strength was calculated with a single-particle model. Since then, many $\beta$-decay and charge-exchange experiments were performed, see e.g. Ref.~\cite{Langanke2021} and references therein, and have motivated the development of more accurate models. 

Two methods arise for computing EC rates at finite temperature. One can determine the rates from each of the initial states in the parent nucleus and compute the Boltzmann-weighted sum of these rates. The other method consists of computing directly temperature-dependent strength functions. The first approach is related to large scale shell model (LSSM) calculations~\cite{Langanke2000,Langanke2003,Oda1994,Tan2020,PhysRevC.83.044619,Suzuki_2016} and the second is related to random-phase approximation (RPA)~\cite{Paar2009,PhysRevC.86.035805}, (relativistic) quasi-particle random phase approximation (QRPA)~\cite{Niu2011,Nabi2004,Ravlic2020} or relativistic time blocking approximation (RTBA)~\cite{Litvinova2021} calculations. Alternatively, one can use hybrid approaches~\cite{Dean1998,Juodagalvis2010,Dzhioev2020}, in which the partial shell occupation numbers at finite temperature are calculated within shell-model Monte-Carlo (SMMC) or Fermi-Dirac parametrizations. Subsequently, these partial occupation numbers are then used as inputs for RPA or QRPA calculations. 

In addition, an analytic approximation of the electron-capture rate as a function of the Q-value was proposed in~\cite{Fuller1985}. The first parametrized version of this approximation~\cite{Langanke2003prl} was fitted to rates on pf-shell nuclei obtained with a hybrid SMMC-RPA approach. Then, for improving the reliability of the extrapolation beyond pf-shell nuclei and far from stability, the parametrization was extended~\cite{Raduta2017} to take into account the effect of the high electron density, temperature, and isospin ratio. 

So far, no EC rate tables from finite-temperature microscopic calculations cover the region of interest for the collapse phase of CCSN, along $N$=50 near $^{78}$Ni, here referred to as the “diamond region”. The first extensive calculations in this region were performed with a hybrid model~\cite{Langanke2003}, but only for a subset of the nuclei of interest, or with a QRPA model~\cite{Titus2019} for all nuclei in the diamond region but without considering temperature-dependent effects. Recently, few finite-temperature calculations~\cite{Dzhioev2020,Litvinova2021} were performed on selected nuclei in the region of interest. These studies show the importance of including higher-order correlations and thermal excitations for explaining the unblocking of the GT+ strength in the nuclei near $N=50$, as well as the significant contribution of forbidden transitions to the total electron-capture rate for some $N=50$ nuclei. Furthermore, application of the relativistic FT-QRPA in Ref. \cite{Ravlic2020} has demonstrated the importance of including pairing correlations for temperatures below the critical temperature of pairing collapse, as well as the sensitivity of EC rates to the strength of the isoscalar pairing in the residual interaction. In this work, we will present new finite-temperature EC rates, available in FFN grid format, from two state-of-art finite temperature QRPA calculations covering the whole diamond region (71 nuclei). In order to have a better insight on those results, we will discuss the effect of the detailed nuclear structure on the electron capture rate, using a new shell model calculation for $^{86}$Kr. The new results presented here will help quantifying the impact of the thermally induced weak-transitions, GT+ but also first-forbidden transitions, on the dynamics of the CCSN.

This paper is structured as follows. In Sec.~\ref{sec:electron_capture}, the formalism used to make new finite-temperature EC rates libraries from two non-relativistic and relativistic finite-temperature QRPA models, are presented. The two models are presented in the following sections ~\ref{sec:nonrel_qrpa} and~\ref{sec:rel_ftqrpa}, respectively. In Sec.~\ref{sec:SM} we give details about the electron-capture rates results based on large-scale shell model calculations. Then, in Sec.~\ref{sec:comparison}, we compare the temperature dependent electron-capture rates computed from the different formalisms introduced previously. Afterwards, in Sec.~\ref{sec:simulation} the outcomes of CCSN simulations based on the new finite-temperature EC rates libraries are compared. Finally, the main conclusions of this work are outlined in Sec.~\ref{sec:conclusion}.

\section{Electron-capture rates calculated from QRPA strength functions}\label{sec:electron_capture}

In a highly-dense and hot pre-supernova environment atoms are fully ionized, leaving free nuclei immersed in an electron plasma described by a Fermi-Dirac distribution of electrons. In order to derive EC rates within such an environment we follow the formalism developed by Walecka et al. in Refs. \cite{PhysRevC.6.719, walecka1975muon, walecka2004theoretical}. Fermi's golden rule relates the electron-nucleus differential cross section to a transition matrix element through
\begin{equation}\label{eq:fermi}
\frac{d \sigma}{d \Omega} = \frac{1}{(2\pi)^2} V^2 E_\nu^2  \frac{1}{2} \sum \limits_{lept. spin.} \frac{1}{2J_i +1} \sum \limits_{M_i M_f} | \langle F | \hat{H}_W | I \rangle |^2,
\end{equation}
where $V$ is the normalization volume, $E_\nu$ is the (massless) neutrino energy, $\ket{I}$ denotes the initial state of the nucleus-electron system, and $\ket{F}$ is the final state (which includes the daughter nucleus and emitted neutrino). The nuclear state has angular momentum $J_i$ and projection $M_i$ before decay and respectively $J_f$ and $M_f$ after decay. We assume the current-current form of the weak interaction Hamiltonian
\begin{equation}\label{eq:hw}
\hat{H}_W = - \frac{G}{\sqrt{2}} \int d^3 \boldsymbol{r} j_\mu^{lept.}(\boldsymbol{r}) \hat{\mathcal{J}}^\mu(\boldsymbol{r}),
\end{equation}
where $G$ is the Fermi constant, $j_\mu^{lept.}(\boldsymbol{r})$ is the lepton current and $\hat{\mathcal{J}}^\mu(\boldsymbol{r})$ is the hadron current. Coordinate-space vectors are denoted by boldface symbols. Performing the multipole expansion of $\bra{F} \hat{H}_W \ket{I}$ and inserting the result into Fermi's golden rule, while performing the sums over lepton spins we obtain the final expression for EC cross sections which can be found in Refs. \cite{PhysRevC.6.719, walecka1975muon, walecka2004theoretical}. It contains the nuclear matrix elements of charge $\hat{\mathcal{M}}_J$, longitudinal $\hat{\mathcal{L}}_J$, transverse electric $\hat{\mathcal{T}}_J^{el.}$ and transverse magnetic $\hat{\mathcal{T}}_J^{mag.}$ multipole operators. These can be readily evaluated within the FT-QRPA. 

While Section~\ref{sec:rel_ftqrpa} discusses the relativistic treatment of EC rates including first-forbidden contributions, to simplify further discussion we present EC rate expressions assuming allowed Gamow-Teller transitions in the low momentum-transfer approximation. This approach is taken in Section~\ref{sec:nonrel_qrpa}, and corresponds to a non-relativistic reduction of expressions by Walecka et al.~\cite{PhysRevC.6.719, walecka1975muon, walecka2004theoretical}. However, differences between the two approaches are small for electrons with energies of up to 40 MeV as exemplified in Ref.~\cite{PhysRevC.100.025801}. In this limit the weak interaction reduces to the Gamow-Teller operator $\vec{\sigma}\hat{\tau}^\pm$, and we compute the total contribution to the stellar EC decay rates by averaging over initial states and summing over final states the phase space weighted transition strength,
\begin{equation}
    \lambda = \frac{\ln 2}{\kappa} \frac{1}{Z} \sum\limits_{i,f} e^{-\beta E_i} \lvert \bra{f} \vec{\sigma} \hat{\tau}^+ \ket{i} \rvert^2 f(W_0^{(i,f)})
    \,.
\end{equation}
Here $\kappa = 6147$~s, $Z = \sum \limits_i (2J_i+1)e^{-\beta E_i}$ is the partition function, and $\ket{i(f)}$ are the initial (final) nuclear states. The phase space factor is dimensionless, defined in terms of the electron mass,
\begin{equation}\label{eq:ftec_phasespace}
\begin{aligned}
    f(W_0^{(i,f)}) = 
    \int_{W^{(i,f)}_{\text{th}}}^{\infty} &\ 
    p W (W_{0}^{(i,f)}  +  W)^{2}\
    \\& \times F_{0}(Z,W) L_{0} f_e(W)\, dW
    \,,
\end{aligned}
\end{equation}
where $W = E_e/(m_e c^2)$ is the total electron energy, ${p=\sqrt{W^2 - 1}}$ is the electron momentum, and $f_e(W)$ is the electron occupation factor in a Fermi gas,
\begin{equation}
    f_e(W) = \left[1 + \exp( \frac{W - {\mu}/{(m_e c^2)}}{k_b T} )\right]^{-1}
    \,.
\end{equation}
The neutrino momentum is $p_\nu = W_0^{(i,f)} + W$. It depends on the maximum positron energy for a $\beta^+$ decay from parent state $i$ to daughter state $f$,
\begin{equation}
\begin{aligned}
    W_0^{(i,f)} 
    &= (M_{N_i} - M_{N_f} + E^{*}_i - E^{*}_f)/(m_e c^2)
    \\
    &=  1 + (Q_{\beta^+} + E^{* i} - E^{*f})/(m_e c^2)
    \,,
\end{aligned}
\end{equation}
where $Q_{\beta^+}$ is the $\beta^+$ $Q$-value, $M_{N_i}$ ($M_{N_f})$ is the initial (final) nuclear mass, and $E^*_i$ ($E^*_f$) is the excitation energy of the parent (daughter).
The condition that $p_\nu > 0$ defines a threshold energy for the captured electron,
\begin{equation}\label{eq:threshold_energy}
W_{\text{th}}^{(i,f)} =
\begin{cases}
  1 & W_0^{(i,f)}  \ge -1
  \\
  \lvert {W_0^{(i,f)}} \rvert &  W_0^{(i,f)}  < -1
\end{cases}
\,.
\end{equation}
The remaining quantities needed in Eq.~\eqref{eq:ftec_phasespace} are the electron chemical potential $\mu$ (which includes the electron rest mass), and the Fermi function $F_0(Z,W)$ and Coulomb function $L_0$~\cite{Mustonen2014}.

To connect with the FT-QRPA, we use the $Q$-value approximation of Ref.~\cite{Engel1999} for $Q_{\beta^+}$ to express $W_0^{(i,f)}$ as a function of the QRPA energy,
\begin{equation}\label{eq:threshold_energy_qrpa}
    W_0^{i,f} = W_0^{k} \approx 1 + (\lambda_p - \lambda_n - \Delta M_{n-H} - \Omega_k)/(m_e c^2)
    \,.
\end{equation}
$\Delta M_{n-H}$ is the neutron-hydrogen mass difference, and $\lambda_n$ ($\lambda_p$) is the neutron (proton) Fermi energy.
At a given energy, the FT-QRPA strength function $\widetilde{S}_F(\omega)$ approximates the ensemble averaged strength for all transitions with energy difference $E_f - E_i \approx \Omega_k$~\cite{our_theory_paper}, i.e.,
\begin{equation}
\begin{aligned}
    \text{Res}&\left[\frac{\widetilde{S}_F(\omega)}{1 - e^{-\beta \omega}},\ \Omega_k\right] \\
    &\approx \frac{1}{Z} \sum\limits_{i,f} e^{-\beta E_i} \lvert \bra{f} \vec{\sigma} \hat{\tau}^+ \ket{i} \rvert^2 
    \quad\forall\quad
    E_f-E_i \approx \Omega_k
    \,.
\end{aligned}
\end{equation}
The rate can therefore be expressed as a single sum over QRPA energies,
\begin{equation}\label{eq:EC_rate}
    \lambda = \frac{\ln 2}{\kappa} \sum\limits_k \text{Res}\left[\frac{\widetilde{S}_F(\omega)}{1 - e^{-\beta \omega}},\ \Omega_k\right]\, f(W_0^k)
    \,.
\end{equation}

While the range of relevant energies is in principle from $-\infty$ to $+\infty$ --- the lower bound to account for de-excitations with infinitely large $Q$-values, and the upper bound to account for capture of infinitely energetic electrons in the Fermi gas --- the phase space function dies off rapidly for larger energies and the exponential prefactor in Eq.~\eqref{eq:ft_prefactor} rapidly dies to zero at negative energies. Thus, in practice a finite energy range can be chosen for a given temperature and chemical potential $\mu$.

\section{Non-relativistic Skyrme FT-QRPA calculation}\label{sec:nonrel_qrpa}
\subsection{Computational method}

In this section we discuss the details of the non-relativistic, axially-deformed Skyrme FT-QRPA calculation with the charge-changing finite amplitude method (FAM)~\cite{Mustonen2014}. We use the SKO' Skyrme functional optimized for the global calculations in Refs.~\cite{Ney2020,Mustonen2016}. This functional was fit with an effective axial vector coupling of $g_A = 1.0$, and was also used for the electron capture calculations in Ref.~\cite{Titus2019}. In that work, the FAM was used to compute Gamow-Teller strength functions at zero temperature with an artificial Lorentzian width of $0.25$~MeV. Odd nuclei were treated in the equal filling approximation (EFA)~\cite{Perez-Martin2008,Shafer2016,Ney2020}. These strength functions were weighted with the temperature- and density-dependent phase space function, Eq.~\eqref{eq:ftec_phasespace}, to estimate stellar electron capture rates. Here, we extend the work of Ref.~\cite{Titus2019} by accounting for the temperature dependence of the Gamow-Teller strength with the FT-QRPA.

We make several adjustments to the calculations performed in Ref.~\cite{Titus2019} to accommodate finite temperature. Odd nuclei in the present work are treated by constraining the finite-temperature Hartree-Fock-Bogoliubov (FT-HFB)~\cite{GOODMAN198130} ensembles to have the desired odd particle number on average. We cannot use the equal filling approximation because it is based on a statistical ensemble formalism, and there is currently no method to treat the EFA ensemble simultaneously with the finite temperature ensemble. Additionally, rather than using strength functions with an artificial Lorentzian width, we compute EC rates using the complex contour integration method described in Ref.~\cite{Mustonen2014}. Although we do not gain any information about the strength distribution using this method, it is significantly less computationally expensive. Moreover, the contour integration method eliminates the artificial width from the calculations completely, providing rates that are comparable to those computed with the matrix form of the FT-QRPA.

\subsection{The finite amplitude method}\label{sec:finite_amplitude_method}

An extension of the FAM to statistical ensembles was discussed in Refs.~\cite{Ney2020,Shafer2016_Thesis} in the context of the EFA. Here we present a similar discussion for the finite-temperature ensemble. The FT-QRPA is equivalent to the free linear response of a finite-temperature HFB ensemble. The corresponding linear response equations were derived in Ref.~\cite{Sommermann1983} and can be written as
\begin{equation}\label{eq:ft_linear_response}
\begin{gathered}
  \Big[ \widetilde{\mathcal{S}} - \omega M \Big] \delta \widetilde{R}(\omega) =  -T \mathcal{F}(\omega)
  \\
  \widetilde{\mathcal{S}} \equiv T\mathcal{H} + \mathcal{E}
  ,\quad
  \delta \widetilde{R} \equiv T \delta R
  \,.
\end{gathered}
\end{equation}
In the notation of Ref.~\cite{Sommermann1983}, we have defined matrices in an extended $4\times4$ supermatrix space and in a two-quasiparticle basis,
\begin{equation}
\begin{aligned}
    T_{\alpha \beta, \gamma\delta}
    % &=\text{diag}[(f_\beta - f_\alpha), (1 -f_\alpha - f_\beta), (1 -f_\alpha - f_\beta), (f_\beta - f_\alpha)] \delta_{\gamma \delta}
    &=\text{diag}[ f_{\beta \alpha}^-, (1 - f_{\alpha \beta}^+), (1 - f_{\alpha \beta}^+),  f_{\beta \alpha}^-] \delta_{\gamma \delta}
\\
    \mathcal{E}_{\alpha \beta, \gamma\delta}
    % &=\text{diag}[(E_\alpha - E_\beta),  (E_\alpha + E_\beta), (E_\alpha + E_\beta), (E_\alpha - E_\beta)] \delta_{\gamma \delta}
    &=\text{diag}[E_{\alpha \beta}^-,  E_{\alpha \beta}^+, E_{\alpha \beta}^+, E_{\alpha \beta}^-] \delta_{\gamma \delta}
\\
    M_{\alpha\beta,\gamma\delta}
    &= \text{diag}[\phantom{-}1,\phantom{-}1,-1,-1] \delta_{\alpha \beta, \gamma \delta}
\\
    \mathcal{H}_{\alpha \beta, \gamma\delta}
    &= \frac{\partial H_{\alpha \beta}}{\partial R_{\gamma\delta}}
    \,.
\end{aligned}
\end{equation}
Matrix elements of $T$ and $\mathcal{E}$ depend on the quasiparticle occupations $f_k = \left[ 1 + \exp \left(E_k/k_B T \right) \right]^{-1},
$ and energies $E_k$ for two quasiparticles, and our shorthand notation means, e.g.,  $E^{\pm}_{\alpha \beta} \equiv E_{\beta} \pm E_{\alpha}$.
The matrix $\mathcal{H}$ represents the residual interaction, where expressions for its sub-matrices are given in Appendix~B of Ref.~\cite{Sommermann1983}.
Finally, the vectors in Eq.~\eqref{eq:ft_linear_response} are the density response, $\delta {R}_{\alpha\beta}(\omega) = (P_{\alpha\beta}, X_{\alpha\beta}, Y_{\alpha\beta}, Q_{\alpha\beta})$, and the external field, $\mathcal{F}_{\alpha\beta}(\omega) = (F^{11}_{\alpha\beta}, F^{20}_{\alpha\beta}, F^{02}_{\alpha\beta}, F^{\bar{11}}_{\alpha\beta})$.

The FAM avoids the expensive construction of the residual interaction matrix by computing the perturbation of the Hamiltonian directly with a finite difference,
\begin{equation}
\begin{aligned}
    \delta \widetilde{H}(\omega) 
    &=
    % \begin{pmatrix} 
    (
    \delta \widetilde{H}^{11}, %\\ 
    \delta \widetilde{H}^{20}, %\\
    \delta \widetilde{H}^{02}, %\\
    \delta \widetilde{H}^{\bar{11}}
    )
    % \end{pmatrix}
    \\
    &= \frac{\partial H}{\partial R} \Bigg\rvert_{R=\widetilde{R}_0} \delta \widetilde{R}(\omega) 
    \\&
    =
    \lim_{\eta \rightarrow 0} \frac{1}{\eta} \left[H[ \widetilde{R}_0 + \eta \delta \widetilde{R}(\omega)] - H[ \widetilde{R}_0] \right]
    \,,
\end{aligned}
\end{equation}
where $\widetilde{R}_0$ is the FT-HFB solution for the generalized density.
Equation~\eqref{eq:ft_linear_response} can then be rearranged to give the FT-FAM equations,
\begin{equation}\label{eq:ft_FAM}
    \left[\mathcal{E} - \omega M\right] \delta \widetilde{R}(\omega) = T \left[\delta \widetilde{H}(\omega) + \mathcal{F}(\omega)\right]
    \,.
\end{equation}
In the charge-changing case, for Skyrme functionals without proton-neutron mixing we can directly evaluate the Hamiltonian perturbation with the perturbed density, i.e., $\delta \widetilde{H} = H[\delta \widetilde{R}]$. Once we have solved the FAM equations, the strength function can be computed from the density response with,
\begin{equation}\label{eq:ft_fam_strength}
\begin{aligned}
    \widetilde{S}_F(\omega) 
    &= 
    \mathcal{F}^\dagger \delta \widetilde{R}(\omega)
    \\&
    = 
    \sum\limits_{k \pm >0} \left[ 
    \frac{
    \big\lvert 
        \big\langle \big[ 
            \Gamma^{k}, \hat{F}
        \big] \big\rangle
    \big\rvert^2}
         {\omega - \Omega_k} -
    \frac{
    \big\lvert
        \big\langle \big[ 
            \Gamma^{k}, \hat{F}^\dagger
        \big] \big\rangle    
    \big\rvert^2}
        {\omega + \Omega_k} \right]
        \,,
\end{aligned}
\end{equation}
where the sum is over FT-QRPA modes with positive norm, while $\Gamma^{k\dagger}$ is the FT-QRPA phonon creation operator defined in Ref. \cite{SOMMERMANN1983163}. To avoid the poles in the strength function, the FAM computes it at complex energies, $\omega_\gamma = \omega + i \gamma$, which smears the poles with Lorentzians of half-width $\gamma$.

The residues of this function contain ensemble averaged transition amplitudes for excitations (located at $\omega>0$) and de-excitations (located at $\omega<0$) of the forward and reverse processes. We can extract the physical transition strength distribution for the forward process, $dB/d\omega$, from Eq.~\eqref{eq:ft_fam_strength} using the relation~\cite{our_theory_paper},
\begin{equation}\label{eq:ft_prefactor}
    \frac{dB}{d\omega} =  -\frac{1}{\pi} \Im\left[ \frac{\widetilde{S}_F(\omega)}{1 - e^{-\beta \omega}}\right]
    \,.
\end{equation}
Unlike the zero-temperature case, Eq.~\eqref{eq:ft_prefactor} is defined for both positive and negative energies (but undefined at ${\omega=0}$ due to the pole from the exponential factor).

\subsection{Phase space integrals}

According to Eq.~\eqref{eq:EC_rate}, the rate depends on two quantities: the transition matrix elements and the phase space integrals. To fully take into account Coulomb effects, we compute the phase space integrals in Eq.~\eqref{eq:ftec_phasespace} numerically. In contrast, the analytic integral used in Refs.~\cite{Reyes2006,Titus2019} requires a more approximate treatment of the Fermi function.

The Fermi-Dirac distribution $f_e(W)$ causes the phase space integrand to change behavior around $W = \mu/m_e c^2$. Below this energy, it behaves mostly like an increasing polynomial, while above $\mu/m_e c^2$ it is mostly a decaying exponential.
This suggests at least two quadratures are necessary to get an accurate result~\cite{Aparicio1998}. We therefore use Gauss-Legendre quadrature for energies below $\mu + \epsilon$ and Gauss-Laguerre for energies above this. $\epsilon$ is a small positive quantity that improves the quadrature's performance at low temperatures where the exponential decay is very steep. We use a value of $\epsilon = 0.1$~MeV and an 80 point grid for both quadratures, which provides very reliable results.

To carry out the integration, we also require the chemical potential, which is a function of the temperature and density. We compute the chemical potential ``on the fly'' by inverting the condition of charge-neutrality in the stellar medium~\cite{Timmes1999},
\begin{equation}\label{eq:rhoye}
\begin{aligned}
    Y_e \rho =& \frac{\sqrt{2}}{\pi^2 N_A} \left( \frac{m_e c^2}{\hbar c} \right)^3 
    % \\& \times \int\limits_0^\infty (S_{e^-}(W,T,\mu) - S_{e^+}(W,T,-\mu-2 m_e c^2)) p^2\, d p
    \beta^{3/2} \Big\{ \Big[ F_{1/2}(\eta, \beta') + \beta'\, F_{3/2}(\eta, \beta') \Big]
    \\&
    -\Big[F_{1/2}(-\eta-2/\beta', \beta') + \beta'\, F_{3/2}(-\eta-2/\beta', \beta') \Big] \Big\}
    \,.
\end{aligned}
\end{equation}
Here $\eta = \mu/(k_b T)$, $\beta' = k_b T/(m_e c^2)$, and $F_k(\eta, \beta)$ is a generalized Fermi-Dirac integral that we compute using the method developed in Ref.~\cite{Aparicio1998}.

\subsection{Contour integration}

As for the transition matrix elements, the FAM can compute them with Eqs.~\eqref{eq:ft_fam_strength} and~\eqref{eq:ft_prefactor}. The complex contour integration method converts the sum over residues in Eq.~\eqref{eq:EC_rate} to a complex contour integration,
\begin{equation}\label{eq:cci_EC_rate}
    \lambda = \frac{\ln 2}{\kappa} \frac{1}{2\pi i} \oint_{C} d \omega\ \frac{\widetilde{S}_F(\omega)}{1 - e^{-\beta \omega}}\ f(\omega)
    \,,
\end{equation}
where we take the contour $C$ to be a circle centered on the real axis.

As demonstrated in Ref.~\cite{our_theory_paper}, treating finite-temperatures with this method introduces several numerical challenges. We use the same procedure as in that work to deal with the poles coming from the exponential prefactor in Eq.~\eqref{eq:cci_EC_rate}. When we need to integrate strength at positive and negative energies, we use two circular contours that pass through $\omega=0$. Each contour picks up half the residue of the spurious pole at $\omega=0$ coming from the prefactor. We therefore also perform a contour integration around just this pole to subtract its contribution. For low temperatures and large contours, poles from the prefactor along the imaginary axis get close to the edge of the contours and cause the integrals to be inaccurate. In such cases, we deform the contours into ellipses to keep them sufficiently far away from the poles on the imaginary axis. For temperatures below 1.0~GK, we neglect the exponential prefactor and strength from de-excitations altogether.

For stellar EC rates, several other numerical challenges arise. Just as in the zero-temperature case, the stellar EC phase space function (Eq.~\eqref{eq:ftec_phasespace}) is not complex analytic and must be approximated by a function that we can evaluate in the complex plane~\cite{Mustonen2014}. However, the phase space integrals exhibit two problematic features.  First, similarly to their integrands, the $f(\omega)$ change behavior when the threshold energy equals the chemical potential, i.e., when ${W^{i,f}_{0}(\omega)} = - \mu/(m_e c^2)$. Second, above this energy the exponential decay causes $f(\omega)$ to approach zero very rapidly. A function with these properties is not able to be approximated well by a simple analytic function, like a polynomial or rational function.

To address the former issue, for a given temperature and density, if the $\omega$ corresponding to ${W^{i,f}_{0}(\omega)} = -\mu/(m_e c^2)$ lies inside the contour bounds, we split the contour in two at that energy. The contour at smaller QRPA energies uses a $6^{\text{th}}$-order polynomial fit\footnote{Assuming the Fermi function is a constant, the indefinite integral of Eq.~\eqref{eq:ftec_phasespace} at energies below the Fermi energy contain $5^{\text{th}}$ order polynomials~\cite{Reyes2006}.} to the phase space integrals, while the one at higher energies uses an exponential fit. As for the latter issue, if the phase space integral falls below machine precision at an energy less than the upper bound of a contour, we shrink the contour to exclude energies above this value. This avoids poorly conditioned exponential fits, which can be extremely oscillatory in the complex plane.

\begin{figure}
\caption{\label{fig:contours} Schematic representations of (a) least and (b) most computationally expensive contour integrations as discussed in the main text. Poles on the real and imaginary axes are black markers, with circles representing poles from the exponential prefactor in Eq.~\eqref{eq:ft_prefactor} and crosses poles from the QRPA strength function.}
\centering
\includegraphics[width=.45\columnwidth]{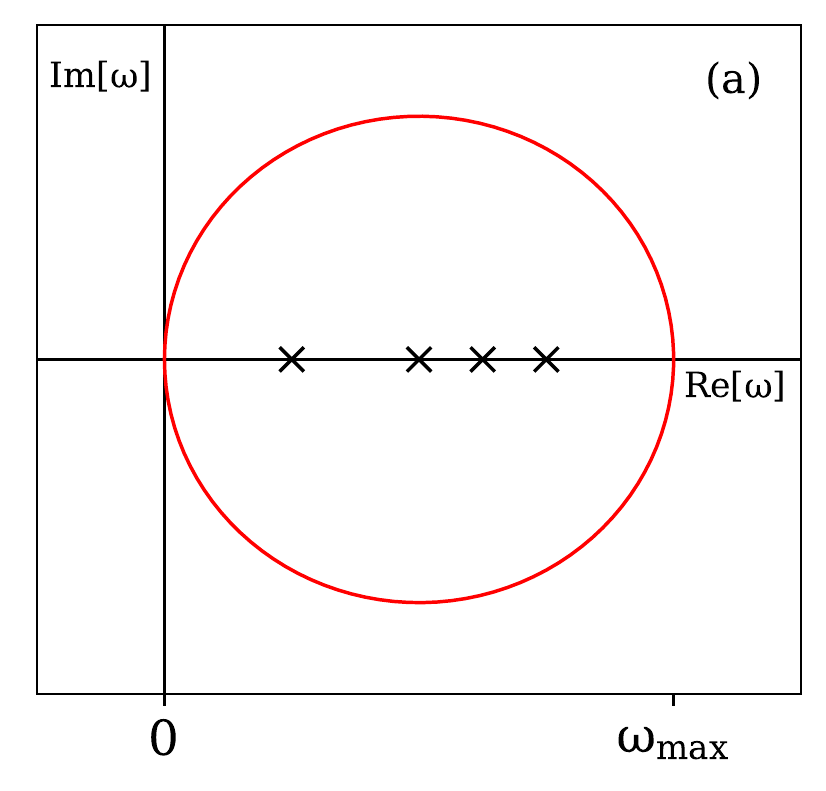}
\includegraphics[width=.45\columnwidth]{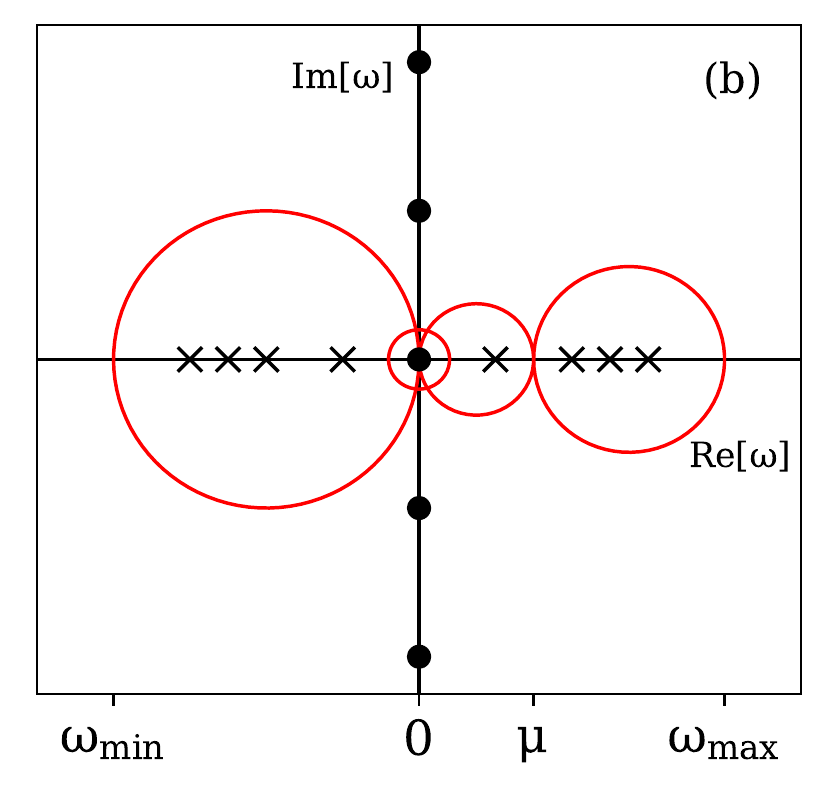}
\end{figure}

For the rates computed in this work, we considered QRPA energies from $-30~\text{MeV}$ to $+30~\text{MeV}$. We use a minimum energy cutoff defined as the energy at which the exponential prefactor becomes smaller than $10^{-20}$ for $T > 1.0$~GK, or zero for $T\leq1.0$~GK. For a maximum energy cutoff, we use the energy at which the phase space function becomes smaller than machine precision for the given $\mu$. If either cutoff is less than the $\pm 30$~MeV bounds, we reduce the energy range accordingly. 

We compute the Gamow-Teller contribution to the rates for the 78 nuclei identified in Refs.~\cite{Titus2018,Titus2019} to be important for core-collapse supernovae on the temperature and density grid used in Ref.~\cite{FFN1980}. While the strength function and lower energy bound are the same for a given temperature, the $\mu$ and upper energy bound depend also on the density. Thus, for a given temperature and density, the number of contour integrations can range from one, if $T\leq1.0$~GK and the threshold energy for $\mu$ lies outside the range $0$--$30$~MeV (Fig.~\ref{fig:contours}\hyperref[fig:contours]{(a)}), to four, if $T>1.0$~GK and the threshold energy for $\mu$ falls within the relevant energy range (Fig.~\ref{fig:contours}\hyperref[fig:contours]{(b)}). Of course, many contours for a given strength function will be identical, with only the phase space changing. So, the computational expense for a rate at a fixed temperature and for $N_\rho$ densities is less than $ 4 N_\rho \times $ the cost of a zero temperature calculation, but can be much greater than $1 \times$ that cost. We use an 88 point Gauss-Legendre grid for all contours, with the region of dense points always closest to the imaginary axis to improve the numerical stability.

\section{Relativistic FT-QRPA Calculation}\label{sec:rel_ftqrpa}

\subsection{Ground and excited-state calculations}

Relativistic mean field theory (RMF) can be formulated based on relativistic nuclear energy density functionals (EDFs). A variety of different functionals exists e.g. meson-exchange, point-coupling, non-linear and others \cite{NIKSIC2011519}. Within this work we employ the meson-exchange EDF with momentum-dependent self-energies D3C${}^*$ \cite{PhysRevC.75.024304}. The nucleons are treated as point-particles which interact via the minimal set of mesons: isoscalar-scalar $\sigma$, isoscalar-vector $\omega$ and isovector-vector $\rho$-meson, as well as the electromagnetic (EM) field. Thus the total Lagrangian density can be written as \cite{PhysRevC.71.064301, GAMBHIR1990132}
\begin{equation}
\mathcal{L} = \mathcal{L}_N + \mathcal{L}_m + \mathcal{L}_{int},
\end{equation}
where $\mathcal{L}_N$ denotes the free-nucleon Lagrangian
\begin{equation}
\mathcal{L}_N = \bar{\psi}(i \Gamma_\mu \partial^\mu - \Gamma m)\psi,
\end{equation}
where $m$ is the bare nucleon mass and $\psi$ is the Dirac field. Meson Lagrangian $\mathcal{L}_m$ contains free meson fields together with the EM field
\begin{align}
\begin{split}
\mathcal{L}_m &= \frac{1}{2} \partial_\mu \sigma \partial^\mu \sigma - \frac{1}{2} m_\sigma^2 \sigma^2 - \frac{1}{4} \Omega_{\mu \nu} \Omega^{\mu \nu} + \frac{1}{2} m_\omega^2 \omega_\mu \omega^\mu \\
&- \frac{1}{4} \vec{R}_{\mu \nu} \cdot \vec{R}^{\mu \nu} + \frac{1}{2} m_\rho^2 \vec{\rho}_\mu \cdot \vec{\rho}^\mu - \frac{1}{4} F_{\mu \nu} F^{\mu \nu},
\end{split}
\end{align}
with meson masses $m_\sigma, m_\omega, m_\rho$ and field tensors $\Omega_{\mu \nu}, \vec{R}_{\mu \nu}$ and $F_{\mu \nu}$ defined as
\begin{align}
\begin{split}
\Omega_{\mu \nu} &= \partial_\mu \omega_\nu - \partial_\nu \omega_\mu, \\
\vec{R}_{\mu \nu} &= \partial_\mu \vec{\rho}_\nu - \partial_\nu \vec{\rho}_\mu, \\
F_{\mu \nu} &= \partial_\mu A_\nu - \partial_\nu A_\mu,
\end{split}
\end{align}
corresponding to $\omega$-meson, $\rho$-meson and the EM field. Lastly, $\mathcal{L}_{int}$ is the interaction term
\begin{equation}
\mathcal{L}_{int} = -g_\sigma \Gamma \bar{\psi} \psi \sigma - g_\omega \bar{\psi} \Gamma^\mu \psi \omega_\mu - g_\rho \bar{\psi} \vec{\tau} \Gamma^\mu \psi \vec{\rho}_\mu - e \bar{\psi} \Gamma^\mu \psi A_\mu,
\end{equation}
with couplings $g_\sigma, g_\omega, g_\rho$ and $e$. In the above, arrows over symbols denote vectors in the isospin space, $\vec{\tau}$ being the isospin Pauli matrix. Within standard meson-exchange functionals $\Gamma^\mu$ and $\Gamma$ reduce to usual Dirac matrices $\gamma^\mu$ and the unit matrix. However, within derivative coupling (DC) interactions, like D3C${}^*$, they are defined by \cite{PhysRevC.71.064301}
\begin{align}
\Gamma_\mu &= \gamma^\nu g_{\mu \nu} + \gamma^\nu Y_{\mu \nu} - g_{\mu \nu} Z^\nu, \\
\Gamma &= 1 + \gamma_\mu u_\nu Y^{\mu \nu} - u_\mu Z^\mu,
\end{align} 
with definitions \cite{PhysRevC.71.064301}
\begin{equation}
Y^{\mu \nu} = \frac{\Gamma_V}{m^4}m_\omega^2 \omega^\mu \omega^\nu, \quad Z^\mu = \frac{\Gamma_S}{m^2} \omega^\mu \sigma.
\end{equation}
We note that $\Gamma_V$ and $\Gamma_S$ are additional couplings of DC models not present in usual meson-exchange functionals. Couplings $g_\sigma, g_\omega, g_\rho$ are functions of vector density $\rho_v = \sqrt{j_\mu j^\mu}$ defined by the vector-current density $j^\mu = \bar{\psi} \gamma^\mu \psi$ with the general functional form \cite{NIKSIC2011519,NIKSIC20141808,PhysRevC.71.024312}
\begin{equation}
g_i(\rho_v) = g_i(\rho_0) f_i(x), \quad i=(\sigma,\omega,\rho),
\end{equation}
where $\rho_0$ is the saturation density of symmetric nuclear matter, $x = \rho_v / \rho_0$ and $f_i(x)$ is the function defined in Refs. \cite{NIKSIC20141808,PhysRevC.71.024312}. This density dependence of couplings includes the so-called rearrangement terms in the equation of motion containing derivatives of couplings $g_\sigma, g_\omega$ and $g_\rho$ with respect to the density $\rho_v$. For finite nuclei it is sufficient to consider stationary solutions, meaning that only time-components of four vectors are considered. Furthermore, due to charge conservation only third component of isospin vectors is non-vanishing. Finally, relativistic EDF is defined as
\begin{equation}
E_{RMF} = \int d^3 \boldsymbol{r} \mathcal{H}(\boldsymbol{r}),
\end{equation}
where $\mathcal{H}(\boldsymbol{r})$ is the Hamiltonian density. Within this work, ground-state calculations are performed based on the finite-temperature Hartree Bardeen-Cooper-Schrieffer (FT-HBCS) theory assuming spherical symmetry \cite{GOODMAN198130}. Only isovector ($T=1,S=0$) component of the pairing interaction is included, meaning that no proton-neutron mixing is assumed in the ground-state calculation. The FT-HBCS equations are derived by the minimization of grand-canonical potential $\Omega$ with respect to the density as defined in Ref. \cite{GOODMAN198130}. Assuming nuclei within heat bath of temperature $T$ with chemical potential $\lambda_q$ ($q$ denoting protons or neutrons) the grand-canonical potential is defined as
\begin{equation}
\Omega = E_{RMF} - TS - \lambda_q N_q,
\end{equation}
where $S$ is entropy and $N_q$ the particle number (either proton or neutron). At finite-temperature occupation probability of particular single-particle state is
\begin{equation}
n_k = v_k^2(1-f_k) + u_k^2 f_k,
\end{equation} 
where $v_k, u_k$ are the BCS amplitudes and $f_k$ is the Fermi-Dirac factor defined  in Sec. \ref{sec:finite_amplitude_method}. The pairing gap $\Delta_k$ is obtained self-consistently through the gap equation \cite{GOODMAN198130}
\begin{equation}
\Delta_k = \frac{1}{2} \sum \limits_{k^\prime >0} G_{k k^\prime} \frac{\Delta_{k^\prime}(1-2f_{k^\prime})}{E_{k^\prime}},
\end{equation}
where the monopole pairing force $G_{k k^\prime} = G\delta_{k k^\prime}$ is assumed, while the quasiparticle (q.p.) energies are $E_k = \sqrt{(\varepsilon_k - \lambda_q)^2 + \Delta_k^2}$, $\varepsilon_k$ being the single-particle energies. The isovector pairing constants $G$ are determined by reproducing the pairing gaps obtained from five-point formula \cite{bender2000pairing} for all nuclei considered within this work.

For the calculation of excited states we employ the finite-temperature proton-neutron relativistic QRPA (FT-PNRQRPA) which represents a small amplitude limit [cf. Eq. (\ref{eq:ft_FAM})] of a more general time-dependent Hartree-Fock equation. For the particle-hole (ph) part of the residual interaction only $\rho$-meson and $\pi$-meson terms are present, whereas the $\pi$-meson direct term vanishes at the ground-state level due to parity conservation. To account for the contact part of the nucleon-nucleon interaction, additional zero-range Landau-Migdal term is included of the form \cite{PhysRevC.69.054303}
\begin{equation}
V_{\delta \pi} = g^\prime \left( \frac{f_\pi}{m_\pi} \right)^2 \vec{\tau}_1 \vec{\tau}_2 \boldsymbol{\Sigma}_1 \cdot \boldsymbol{\Sigma}_2 \delta(\boldsymbol{r}_1 - \boldsymbol{r}_2),
\end{equation} 
where standard values are used for the pion-nucleon couplings $\quad f^2_\pi/(4\pi) = 0.08$, $m_\pi = 138.0 \text{ MeV}$, and $\boldsymbol{\Sigma} = \begin{pmatrix}
\boldsymbol{\sigma} & 0 \\
0 & \boldsymbol{\sigma}
\end{pmatrix}$,
$\boldsymbol{\sigma}$ being the Pauli matrix. The parameter $g^\prime = 0.76$ is adjusted to reproduce the experimental excitation energy of Gamow-Teller resonance in ${}^{208}$Pb \cite{PhysRevC.71.064301}. For the particle-particle (pp) part of the residual interaction both isovector ($T=1,S=0$) and isoscalar ($T=0,S=1$) terms contribute. For the isovector pairing we employ the pairing part of the Gogny D1S interaction \cite{PhysRevC.86.064313}, while the isoscalar pairing is formulated as a combination of short-range repulsive Gaussian with a weaker long-range attractive Gaussian \cite{PhysRevC.69.054303}
\begin{equation}
V_{12} = V_0^{is} \sum \limits_{j = 1}^2 g_j e^{-r_{12}^2/\mu_j^2} \prod\limits_{S = 1, T=0},
\end{equation}
where $\prod\limits_{S = 1, T=0}$ denotes the projector on $T = 0, S=1$ states. For the ranges we use $\mu_1 = 1.2$ fm, $\mu_2 = 0.7$ fm, and strengths are set to $g_1 = 1$ and $g_2 = -2$ \cite{PhysRevC.69.054303}. In contrast to the isovector pairing which is constrained by the experimental data at the ground-state level, for the strength of the isoscalar pairing we use the following functional form \cite{Wang_2016, NIU2013172}
\begin{equation}
V_0^{is} = V_L + \frac{V_D}{1+e^{a+b(N-Z)}},
\end{equation}
with parameters $V_L = 153.2$ MeV, $V_D = 8.4$ MeV, $a=6.0$ and $b=-0.8$ adjusted to reproduce best all available experimental half-lives in the range $8 \leq Z \leq 82$ as in Ref. \cite{ravlic2020evolution}.

The FT-PNRQRPA eigenvalue problem can be derived from Eq. (\ref{eq:ft_FAM}) by expanding the perturbed density $\delta \tilde{\mathcal{R}}$ in the configuration space of (quasi)proton-(quasi)neutron basis. Here we omit the details and refer the reader to Refs. \cite{PhysRevC.101.044305, PhysRevC.96.024303, PhysRevC.102.065804} for additional information. We denote the eigenvector corresponding to eigenvalue $\Omega_k$ as $\begin{pmatrix}
P^k & X^k & Y^k & Q^k
\end{pmatrix}^T$. Calculations are symmetric with respect to the isospin projection operator, meaning that they can be split into $\Delta T_z = \pm 1$ component, $\Delta T_z$ denoting the change in isospin projection. The ensemble average appearing in Eq. (\ref{eq:ft_fam_strength}) is evaluated as
\begin{align}
\begin{split}
\langle [ \Gamma^k, \hat{F} ] \rangle &=  \sum \limits_{\pi \nu}P^{k*}_{\pi \nu} F^{11}_{\pi \nu}(f_\nu - f_\pi) + X^{k*}_{\pi \nu} F^{20}_{\pi \nu} (1 - f_\pi - f_{\nu}) \\
&+ Y^{k*}_{\pi \nu} F^{02}_{\pi \nu} ( 1-f_\pi - f_{\nu}) + Q^{k*}_{\pi \nu} F^{\bar{11}}_{\pi \nu}(f_\nu - f_{\pi}),
\end{split}
\end{align}
in the quasiparticle proton-neutron ($\pi - \nu$) basis. Within the FT-HBCS the charge-changing external field operator $\hat{F}$ in $\Delta T_z = -1$ direction has the form
\begin{align}
\begin{split}
&F^{11}_{\pi \nu} = u_\pi u_\nu \langle \pi | \hat{F} | \nu \rangle, \quad 
F^{20}_{\pi \nu} = u_\pi v_\nu \langle \pi | \hat{F} | \nu \rangle, \quad \\
&F^{02}_{\pi \nu} = v_\pi u_\nu \langle \pi | \hat{F} | \nu \rangle, \quad 
F^{\bar{11}}_{\pi \nu} = v_\pi v_\nu \langle \pi | \hat{F} | \nu \rangle ,
\end{split}
\end{align}
where $ \langle \pi | \hat{F} | \nu \rangle$ are the single-(quasi)particle matrix elements. The physical strength distribution $dB / d\omega$ is finally calculated from Eq. (\ref{eq:ft_prefactor}).

Within ground-state calculation, equations of motion are solved by expanding nucleon and meson wave functions in the basis of spherical harmonic oscillator. We are using the following prescription: if $T \leq 10$ GK expansion in 18 oscillator shells for both fermion and boson fields is used while for temperatures $T > 10 $ GK we expand in 20 oscillator shells. We have verified that such approach yields excellent convergence. Radial integrations are discretisized within a spherical box of 20 fm with 24 meshpoints of Gauss-Hermite quadrature. Odd nuclei are treated by constraining neutron (proton) chemical potential $\lambda_{n(p)}$ to odd particle number within the FT-HBCS calculation. This approach was already implemented for calculation of $\beta$-decay half-lives throughout the nuclide chart in Ref. \cite{PhysRevC.93.025805}, yielding reasonable agreement with experimental data. Due to the large number of 2 q.p. states within the FT-PNRQRPA we use two constraints: (i) maximal energy cut-off $E_{cut} = 100$ MeV is set for the sum of q.p. energies of particular pair $E_\pi + E_\nu$ and (ii) states with $|u_\pi v_\nu| < 0.01$ or $|v_\pi u _\nu| < 0.01$ are also excluded from calculations having quite small contribution to matrix elements. With these constraints our FT-PNRQRPA matrix never exceeds dimension of $10000 \times 10000$. Furthermore, we neglect the contribution of antiparticle states, which is a good approximation for charge-exchange transitions \cite{PhysRevC.69.054303}.

\subsection{Calculation of electron capture rates}

The relativistic calculations of EC rates are based on the Walecka formalism as described in Sec. \ref{sec:electron_capture}, evaluated by employing the FT-PNRQRPA for particular total angular momentum and parity $J^\pi$. Both allowed ($0^+, 1^+$) and first-forbidden ($0^-, 1^-, 2^-$) transitions are included in the calculations. The axial-vector couping constant $g_A$ is quenched from its free-nucleon value $g_A = -1.26$ to $g_A = -1.0$ based on previous calculations in Refs. \cite{Ravlic2020, PhysRevC.79.054323} that is also consistent with non relativistic calculations in this work. Finally, EC rates are calculated by folding the EC cross sections with the Fermi-Dirac distribution of electrons
\begin{equation}\label{eq:ec-rate}
\lambda = \frac{(m_e c^2)^3}{\pi^2 \hbar^3} \int \limits_{W_{\mathrm{th}}^k}^\infty p W \sigma(W) f_e(W) dW,
\end{equation}
where the threshold energy $W_{\mathrm{th}}^k$ for the FT-PNRQRPA eigenvalue $k$ with energy $\Omega_k$ is defined in Eqs. (\ref{eq:threshold_energy})-(\ref{eq:threshold_energy_qrpa}). Electron chemical potential $\mu$ is evaluated by inverting Eq. (\ref{eq:rhoye}) which determines the electron Fermi-Dirac factors $f_e(W)$. In order to solve for the EC rate in Eq. (\ref{eq:ec-rate}) we observe that due to Fermi-Dirac function, integrand displays a prominent peak when plotted with respect to the electron energy $E_e = W m_e c^2$. As a first step of the integration we search for the energy of the peak $E_{peak}$ within a predefined interval, with upper limit $E_{max} = \mu + 20 k_BT$, that is large enough to include the peak. The integration array is split into 3 parts. If we define $E_1 = E_{peak}-3k_B T$ and $E_2 = E_{peak}+3k_B T$, they are: (i) $[m_e, E_1\rangle$, (ii) $[\text{min}(E_1,m_e), E_2]$ and (iii) $\langle E_2, E_{max}]$. Numerical integration of EC rates within all 3 intervals is performed with the Gauss-Legendre quadrature. Intervals (i) and (iii) contain 16 mesh-points, while number of mesh-points in interval (ii) is calculated as $|E_2 - E_1|/(0.1k_B T)$. We have verified that above integration mesh yields excellent convergence for required temperatures $T$ and stellar densities $\rho Y_e$ within this work.

\section{Shell-model calculation}%
\label{sec:SM}

Although it is challenging to perform shell-model calculation on many nuclei in the $N=50$ region, it is instructive to compare the results from the QRPA calculations for a specific case. We focus on the case of $^{86}$Kr, for which the GT strength distribution has been measured and compared to calculations at zero~\cite{Titus2019} and finite temperature~\cite{Dzhioev2020}. Our shell-model calculations are performed with the code NUSHELLX~\cite{Brown2014} and the jj45c Hamiltonian, and are based on a $^{78}$Ni core with a model space that includes the orbitals ($0f_{5/2}, 1p_{3/2}, 1p_{1/2}, 0g_{9/2}$) for the protons and ($0g_{7/2}, 1d_{5/2}, 1d_{3/2}, 2s_{1/2}, 0h_{11/2}$) for the neutrons. The jj45c Hamiltonian is described in~\cite{Zamora2019}. The proton-neutron two-body matrix elements were obtained from the CD-Bonn potential as described in~\cite{Dillmann2003}. The proton-proton part of the Hamiltonian is taken from~\cite{Lisetskiy2004}. The neutron single-particle energies were adjusted to reproduce the low-lying states of $^{89}$Sr. To account for temperature-dependent effects, GT transitions from the first 50 initial states for each $J^{\pi}=0,1,2,3,4,5,6,7,8^{+,-}$ in $^{86}$Kr were included, reaching to the first 500 final states for each initial state in $^{86}$Br. These initial states cover excitation energies up to $\approx 17$~MeV, but the results shown here are restricted to states with an excitation energies below 12~MeV, as it was found that contributions to the overall electron-capture rate from states above $\approx 10$~MeV was negligible for all stellar temperatures considered here. The GT strengths for the individual transitions were used to calculate the corresponding EC rates, with the code "ECRATES" previously developed and used in~\cite{Reyes2006,Timmes2000,Gupta2007}. EC rates from different initial states were calculated as follows:
\begin{equation}\label{eq:EC_rate_SM}
\lambda_{i}^{EC} = \frac{\ln 2}{\kappa} P_i \sum_{j} B_{ij}\Phi_{ij}^{EC},
\end{equation}
where the constant $\kappa = 6146\pm6$~s can be determined from super-allowed Fermi transitions. In our case $B_{ij}=B_{ij}(\mathrm{GT}+)$ are the reduced transition probabilities of only the GT+ transitions and are obtained from the NUSHELLX code~\cite{Brown2014} including a quenching of 0.77 for the
Gamow-Teller operator. $\Phi_{ij}^{EC}$ is the phase-space integral as defined in Eq.~\eqref{eq:ftec_phasespace}. For a parent nucleus in thermal equilibrium, at the temperature $1/\beta = k_BT$, where $k_B$ is the Boltzmann constant, the probability of populating an excited state $i$ at the energy $E_i$ is given by,  
\begin{equation}\label{eq:PS_SM}
P_{i} =   \frac{ (2J_i+1)e^{-E_i\beta} }{Z},
\end{equation}
where $Z=\sum_i (2J_i+1)e^{-E_i\beta}$ is the partition function.
As can be seen from Eq.~\eqref{eq:EC_rate_SM}, the EC rate on a given initial state depends on three main factors: i) the GT strength of the individual transitions; ii) the phase-space factor, which depends on the temperature and density of the stellar environments, and on the Q-value for the specific EC transition; and iii) the thermal population of the initial state. It is interesting to investigate the interplay between these three factors to better understand the total EC rate at high stellar densities. 

\begin{figure}
\centering
\includegraphics[width=.85\columnwidth]{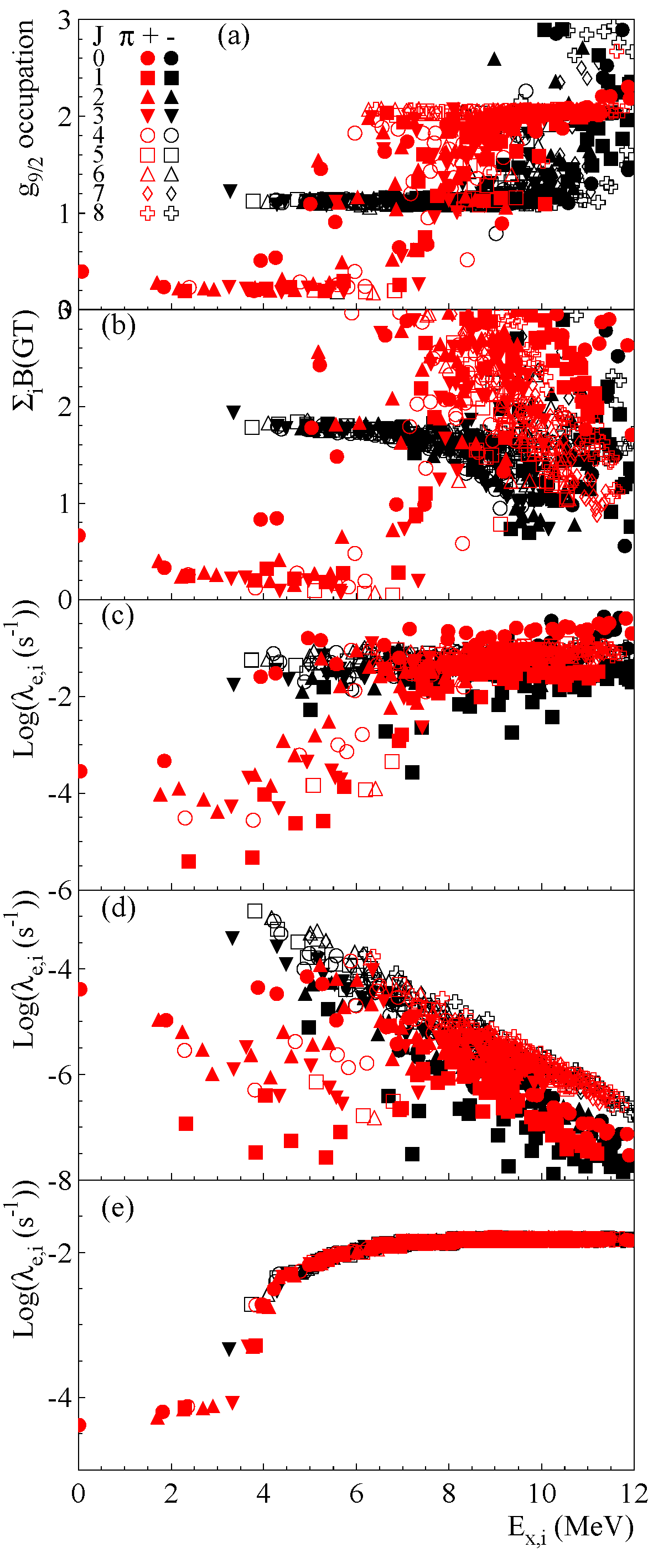}
\caption{\label{fig:86KrSM} Five quantities are plotted against the excitation energy in $^{86}$Kr, based on shell-model and electron-capture rate calculations with the code NUSHELLX~\cite{Brown2014} and ECRATES~\cite{Reyes2006,Timmes2000,Valdez2012}, respectively. a) The average occupation of the $g_{9/2}$ shell in $^{86}$Kr, b) the total GT strength of each initial state in $^{86}$Kr, the logarithm of the electron-capture rates of each initial states in $^{86}$Kr, without c) and with d) weighting with the probabilities of occupying the state i in $^{86}$Kr. e) The logarithm of the cumulative electron-capture rates including thermal population weighting. The contributions from spins of states are represented by different symbols and the parities are distinguished by red (positive) or black (negative) color. The results are obtained at $T = 10$~GK and $\rho Y_e = 10^{9}$~g cm$^{-3}$. These conditions are representative of the start of CCSN and high density burning during SN1a.}
\end{figure}

In the model space considered here, the key factor that determines the GT strength for an individual transition is the filling of the protons in the $g_{9/2}$ shell, as other single-particle contributions to GT excitations are not available. The average population of this shell as a function of excitation energy is shown in Fig.~\ref{fig:86KrSM}\hyperref[fig:86KrSM]{(a)}. Initial states with positive (negative) parity have red (black) labels, and states with different spins have different symbols, as indicated. At low excitation energies, the $g_{9/2}$ shell is only fractionally filled for states with positive parity. At an excitation energy of about 10~MeV, the positive parity states have two protons in the $g_{9/2}$ shell. In the intermediate excitation region, the average filling of the $g_{9/2}$ shell slowly increases. For negative parity states at low excitation energy, the $g_{9/2}$ shell is filled with about one proton. Above 10~MeV, some states have three protons in the $g_{9/2}$ shell, slowly increasing the average population of the $g_{9/2}$ shell.

The filling of the $g_{9/2}$ shell has a profound impact on the GT strengths, as shown in Fig.~\ref{fig:86KrSM}\hyperref[fig:86KrSM]{(b)}. It displays the summed GT strength from each individual initial state to all of its associated final states. Since for the low-lying initial states with positive parity the filling of the $g_{9/2}$ shell is small, the summed GT strengths are mostly significantly below 1. Since the lowest-lying negative parity states (first appearing at $E_{x,i}\approx 3.5$~MeV) have one proton in the $g_{9/2}$ shell, the summed GT strength to all its final states is significantly higher than that for the positive parity states. Above $E_{x,i}\approx 6$~MeV, the spread in summed GT strengths from the initial states increases, as the population of protons in the $g_{9/2}$ shell slowly increases and transitions from positive-parity states generally have higher summed strengths, associated with having two protons in the $g_{9/2}$ shell.

Fig.~\ref{fig:86KrSM}\hyperref[fig:86KrSM]{(c)} shows the EC rates (in logarithmic scale) from each of the initial states, assuming equal population of all initial states. Clearly, the EC rates on the low-lying positive-parity states is much smaller than those on the negative-parity states and the highly excited positive-parity states. This has two causes: i) the lower GT strengths for the low-lying positive parity states as shown in Fig.~\ref{fig:86KrSM}\hyperref[fig:86KrSM]{(a)}, and ii) the favorable Q-value that greatly increases the phase-space factor for transitions from states at high excitation energy. This is due to the fact that for states at high initial excitation energy it is likely that the first final states have low excitation energies in the EC daughter. As the phase-space factor increases exponentially with increasing (more positive) Q-value, the effects of the second cause can have a higher impact than that due to the difference in GT strength. Above $E_{x,i}\approx$5~MeV, we observe that the EC rate becomes almost independent of spin, parity, and excitation energy of the initial state.

Finally, one has to consider the thermal population of the initial state. This is shown in Fig.~\ref{fig:86KrSM}\hyperref[fig:86KrSM]{(d)}, where the data of Fig.~\ref{fig:86KrSM}\hyperref[fig:86KrSM]{(c)} have been weighted by the thermal population factor of Eq.~\eqref{eq:PS_SM}. We note that the EC rates shown here were calculated at $T = 10$~GK and $\rho Y_e = 10^{9}$~g cm$^{-3}$. This corresponds to an environment relatively early in the supernovae collapse phase. At even higher densities and temperatures, the EC rates become even less sensitive to the properties of individual initial and final states, as more initial and final states can contribute. The thermal population factor enhances the contributions from the initial states with the lowest excitation energies. However, it also indicates that the total EC rate is dominated by EC rates on negative-parity states in the initial excitation energy region between 3.5 and 6.0~MeV. The impact of the ($2 J_i +1$) factor is also clear from this figure - contributions from states with higher initial spin are enhanced because of this factor. Fig.~\ref{fig:86KrSM}\hyperref[fig:86KrSM]{(e)}, shows the running sum of the EC rates as function of excitation energy of the initial state. It saturates just above 6~MeV, after the strong contributions from the negative-parity states between 3.5 and 6~MeV. The contributions to the total EC rate from the low-lying positive-parity states only constitutes about 1$\%$ of the total EC rate. 
The model-space considered here is limited, likely causing an underestimation of the EC rates as more complex features are ignored, such as the excitation of protons or neutrons from $0g_{9/2}$ to $0g_{7/2}$. Still, the results indicate that the total EC rate on nuclei in the $N=$50 region depends on an interplay between nuclear structure effects, the EC phase-space factors, and the thermal population of initial states. As a consequence, the total EC rate is not very sensitive to a few nuclear transitions, but rather to the gross nuclear-structure properties in this region.

\section{Comparison of electron-capture rates}%
\label{sec:comparison}

\begin{figure}
\centering
\includegraphics[width=.95\columnwidth]{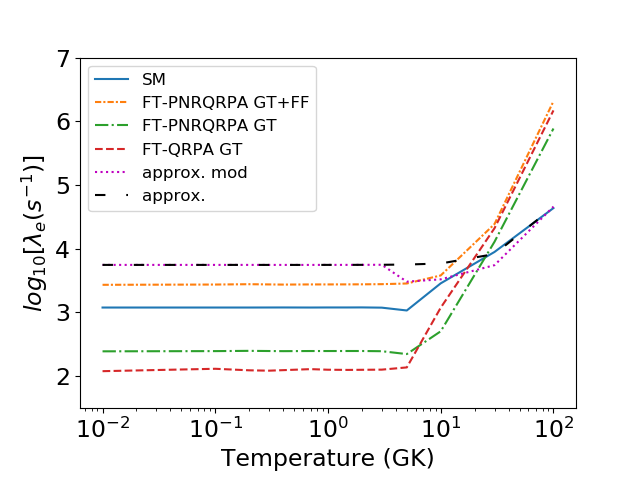}
\caption{\label{fig:86KrCompRates} Electron-capture rate as function of the temperature at $\rho Y_e = 10^{11}$~g.cm$^{-3}$, for the shell-model calculation (SM), and the different temperature-dependent QRPA calculations (FT-QRPA, FT-PNRQRPA GT and FT-PNRQRPA GT+FF including GT and first-forbidden transitions) of this work, as well as for the approximation from~\cite{Langanke2003} and the third version of the modified approximation from~\cite{Raduta2017}.}
\end{figure}

The comparison between our new results for the electron-capture rates on $^{86}$Kr is shown in Fig.~\ref{fig:86KrCompRates},  which shows the EC rate at a density of $\rho Y_e = 10^{11}$~g.cm$^{-3}$ as a function of stellar temperature. Since the first-forbidden contributions are not included in the shell-model (SM) calculations, one can compare the SM to the FT-QRPA and the FT-PNRQRPA GT results. At temperatures below $T \approx 15$~GK the SM rates are higher than the two QRPA calculations. Above this temperature, the opposite is the case. As already discussed in Ref.~\cite{Dzhioev2010}, the QRPA calculations are more sensitive to the effects of increased temperatures than the SM calculations. At low temperatures, the GT strength distribution spread out to higher excitation energies in the QRPA calculations than in the SM calculation, resulting in a lower EC rate. As the temperature increases, GT strengths at low excitation energies are enhanced in the QRPA calculations, leading to a rapid rise in EC rates. This is related with two main effects: (i) vanishing of pairing correlations with increasing temperature, and (ii) thermal unblocking, which allows transitions to previously blocked q.p. states, as demonstrated in Ref. \cite{Dzhioev2020}. On the other hand, at high temperatures, the restrictions to the model space in the SM calculations likely lead to an underestimation of the the EC rates. 
By comparing the FT-PNRQRPA GT and FT-PNRQRPA GT+FF calculations, it is clear that the contributions from the FF transitions are significant. The impact is strongest at temperatures below 30~GK because the GT transitions are strongly Pauli-blocked~\cite{Titus2019}. At higher temperatures, the Pauli blocking is reduced and the contributions to the total EC rate from GT and FF transitions become comparable.
\begin{figure}
\centering
\includegraphics[width=.95\columnwidth]{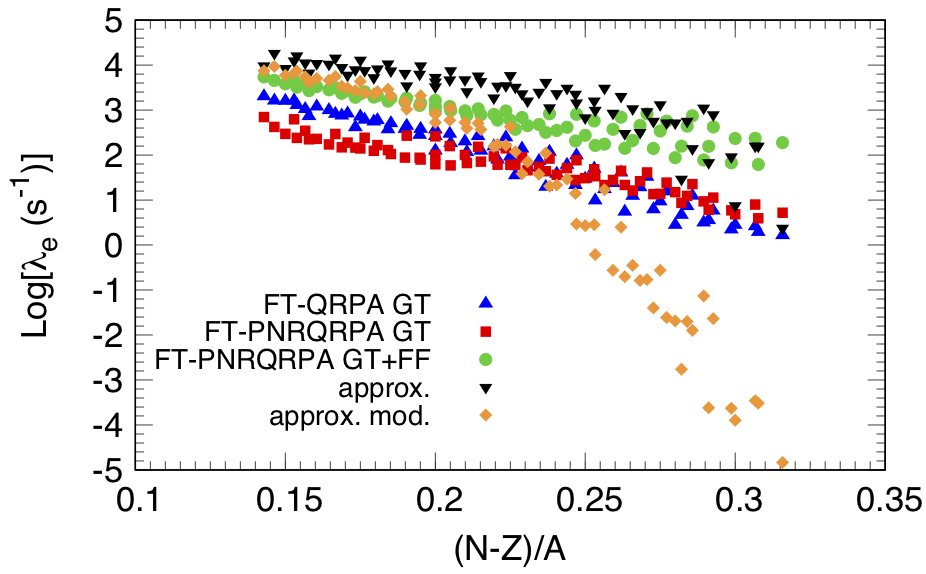}
\caption{\label{fig:ratesI} Electron-capture rate as function of the isospin asymmetry $(N-Z)/A$ at $T=10$~GK and $\rho Y_e = 10^{11}$~g.cm$^{-3}$. Results from various electron-capture rate prescriptions are compared: the shell-model calculation (SM), and the different temperature-dependent QRPA calculations (FT-QRPA, FT-PNRQRPA GT and FT-PNRQRPA GT+FF including GT and first-forbidden transitions) of this work, as well as for the approximation from~\cite{Langanke2003} and the third version of the modified approximation from~\cite{Raduta2017}.}
\end{figure}
Our results are in relatively good agreement with TQRPA results in Ref.~\cite{Dzhioev2020} for which, at $T = 10$~GK and $\rho Y_e = 10^{11}$~g.cm$^{-3}$, the EC rates with all transitions included approach $10^4$~s$^{-1}$ as the FT-PNRQRPA with $3.8\times10^3$~s$^{-1}$. Moreover, the relative contribution of the first-forbidden (FF) transitions to the EC rates $\lambda^{\rm FF}/\lambda = 0.87$ is reasonably close to 0.75 obtained with Skyrme-SkO'-TQRPA in~\cite{Dzhioev2020}. That the results from different sets of calculations are comparable gives confidence that the main nuclear structure features are covered in the calculations.
One may remark that the "approx. mod." curve in Fig.~\ref{fig:86KrCompRates} follows the original approximation~\cite{Langanke2003} bellow $T = 5$~GK for $\rho Y_e = 10^{11}$~g/cm$^{-3}$. Because these conditions correspond to the limits for which the parametrization of the average GT transition energy of Ref.~\cite{Raduta2017} hold, we choose to follow the original parametrization~\cite{Langanke2003} outside of these limits. Above $T\approx 10$~GK, the shell model EC rates and the predictions from the approximation of \cite{Langanke2003} and \cite{Raduta2017}  converge.
\begin{figure}
\centering
\includegraphics[width=.75\columnwidth]{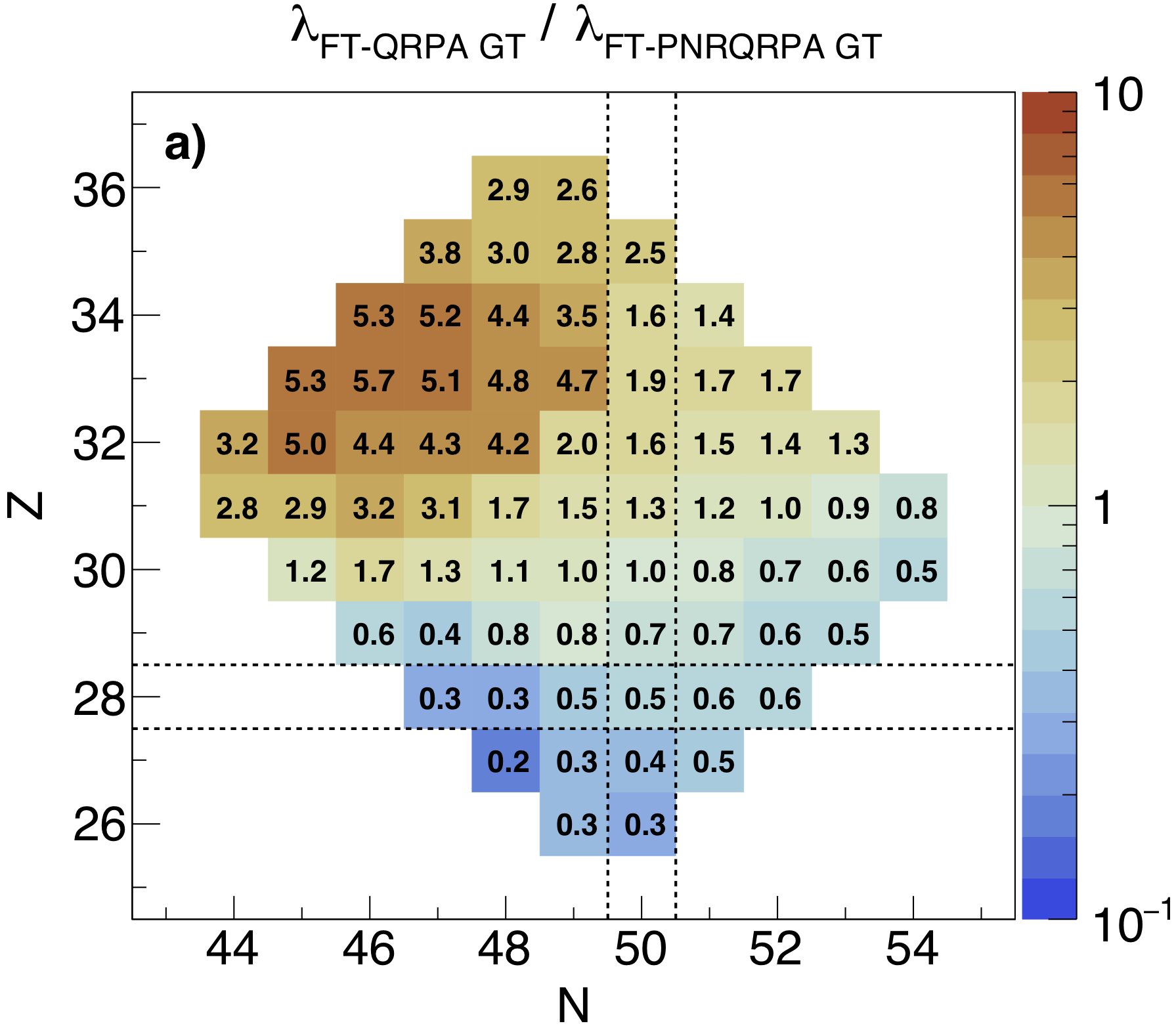}
\includegraphics[width=.75\columnwidth]{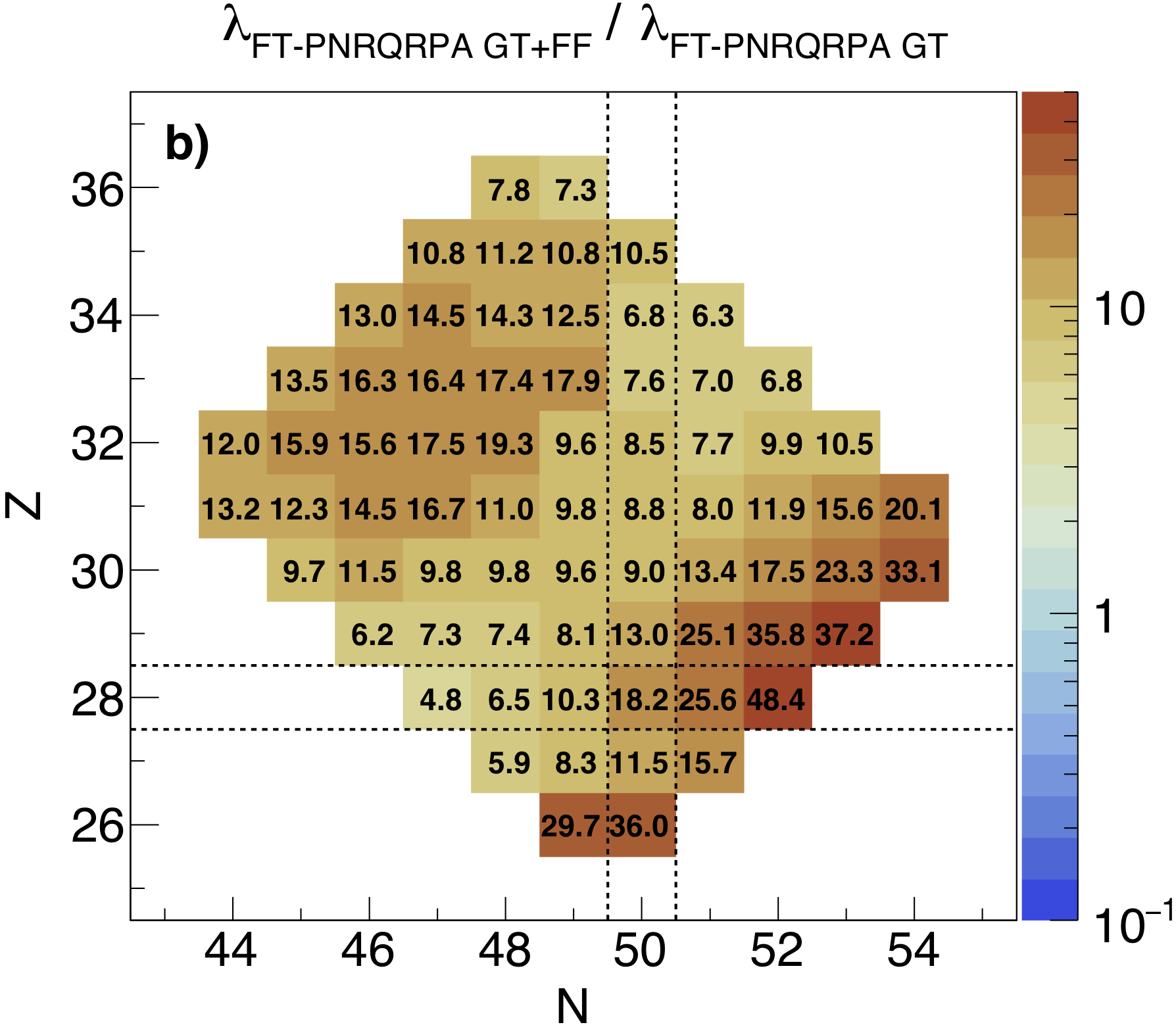}
\caption{\label{fig:compRates2d} Region of the nuclear chart with nuclei dominating the electron-capture rate during core-collapse supernovae, as defined in~\cite{Sullivan2015}. The dashed lines distinguish the shell closures $Z=28$ and $N=50$. The color scale represent a ratio of electron-capture rates from different prescriptions, in (a) the FT-QRPA, over the FT-PNRQRPA, with GT transitions only, in (b) the FT-PNRQRPA with GT over the FT-PNRQRPA with GT and first-forbidden transitions. The rates are obtained at $T=10$~GK and $\rho Y_e = 10^{11}$~g.cm$^{-3}$.}
\end{figure}

Figs.~\ref{fig:ratesI}~and~\ref{fig:compRates2d} illustrate comparisons of the EC rates of all the nuclei in the region of interest for CCSN, at $T = 10$~GK and $\rho Y_e = 10^{11}$~g.cm$^{-3}$. In Fig.~\ref{fig:ratesI}, the EC rates are represented as function of the isospin asymmetry $(N-Z)/A$. Overall, the EC rates from the original approximation~\cite{Langanke2003} are higher than those from the microscopic calculations, except the FT-PNRQRPA GT+FF for a few neutron-rich nuclei ($(N-Z)/A\gtrsim$0.25). The modified approximation (third parametrization in~\cite{Raduta2017}) is the closest to the EC rates of FT-PNRQRPA GT+FF for $(N-Z)/A\lesssim$0.20, but strongly decreases for more neutron-rich nuclei. This can be explained by the refined parametrization of the average GT transition energy in~\cite{Raduta2017}, which increases linearly with $(N-Z)/A$. The new parametrization has been introduced to better fit the EC rates of nuclei with low Q-value. The reference rates of Ref.~\cite{Langanke2001} used in Ref.~\cite{Raduta2017} are obtained with large-scale shell-model calculations of pf-shell nuclei ($45<A<65$), considering only few initial states (4 to 12) and without forbidden transitions. These assumptions can result in underestimating the EC rates of neutron-rich nuclei at finite temperature. 

The agreement between the FT-QRPA and the FT-PNRQRPA GT-only calculations is relatively good especially around $(N-Z)/A=0.24$. A more detailed comparison between these two rate sets is shown in Fig.~\ref{fig:compRates2d}\hyperref[fig:compRates2d]{(a)}, which shows the ratio between the two sets as a function of neutron and proton number. This ratio  varies between 0.2 and 5.7. The EC rates obtained with FT-QRPA model dominate around $^{79}$As ($Z=33, N=36$), whereas the FT-PNRQRPA model predicts higher EC rates for the nuclei around $^{75}$Co ($Z=27, N=48$) and for the most neutron-rich nuclei in general. Such differences can be attributed to the systematic model dependence. The FT-QRPA calculations are performed with the non relativistic EDF with Skyrme SkO' interaction, while the FT-PNRQRPA employs the relativistic derivative coupling EDF. Furthermore, in the non relativistic FT-QRPA calculations axial-symmetry is assumed while the FT-PNRQRPA assumes spherical symmetry. Although a shape-phase transition is expected from deformed to a spherical state at high temperatures \cite{PhysRevC.96.054308,PhysRevC.97.054302}, deformation can persist at $T = 10$ GK, which leads to differences between two sets of EC rates. In Fig.~\ref{fig:compRates2d}\hyperref[fig:compRates2d]{(b)}, the ratio between rates from the FT-PNRQRPA GT-only calculation over the FT-PNRQRPA GT+FF calculation is shown. The ratio averages around 10, but for the most neutron-rich nuclei below $Z=31$ the importance of the first-forbidden transitions increases because Pauli-blocking effects for the GT transitions are strongest in this region.
\begin{figure}
\centering
\includegraphics[width=.75\columnwidth]{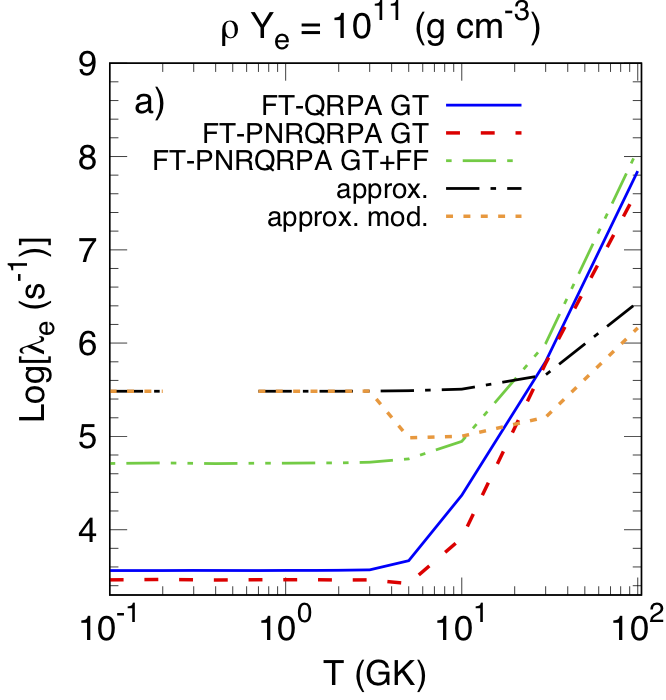}
\includegraphics[width=.75\columnwidth]{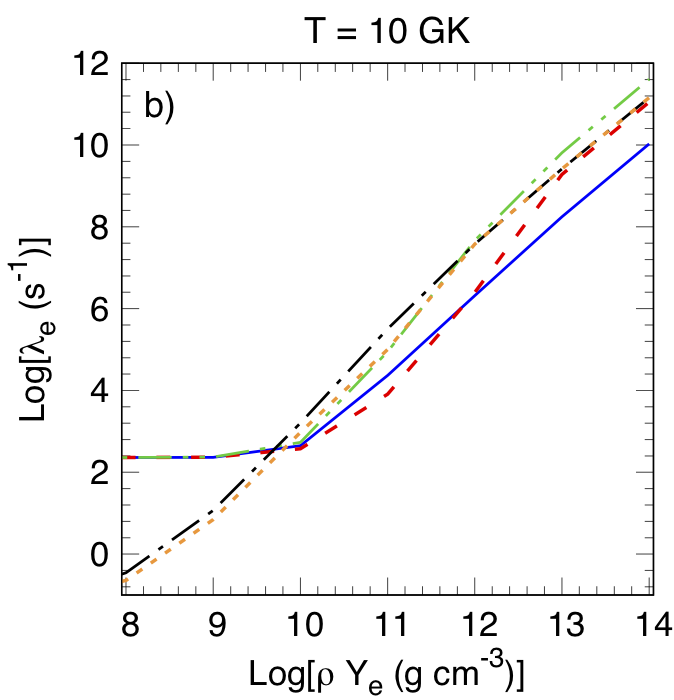}
\caption{\label{fig:compTotalRate} Electron-capture rate as function of the temperature ($T$) for $\rho Y_e = 10^{11}$~g.cm$^{-3}$ (a) and as function of the density $\rho Y_e$ for $T=10$~GK (b). Results from various electron-capture rate prescriptions are compared: the different temperature-dependent QRPA calculations (FT-QRPA, FT-PNRQRPA GT and FT-PNRQRPA GT+FF including GT and first-forbidden transitions) of this work, as well as for the approximation from~\cite{Langanke2003} and the third version of the modified approximation from~\cite{Raduta2017}.}
\end{figure}

Furthermore, in Fig.~\ref{fig:compTotalRate}, we compare the summed EC rates of the nuclei in the diamond region for the different models, as function of the temperature and the density. Note that the EC rates in Fig.~\ref{fig:compTotalRate} are not weighted by the actual populations in the stellar medium, but still the unweighted sum gives a rough understanding of how the EC models may affect the CCSN scenario. As observed with the comparison of the individual EC rates, the FT-QRPA and the FT-PNRQRPA GT calculations agree well, the largest difference is seen in Fig.~\ref{fig:compTotalRate}\hyperref[fig:compTotalRate]{(b)} for densities $\rho Y_e \gtrsim 5 \times 10^{12}$~g/cm$^{-3}$. The relative contribution of the first-forbidden transitions in the FT-PNRQRPA calculations is larger at low temperature ($T \lesssim 10$~GK for $\rho Y_e = 10^{11}$~g.cm$^{-3}$) and high density ( $\rho Y_e \gtrsim 10^{10}$~g/cm$^{-3}$ for $T=10$~GK). As already mentioned, agreement for temperatures above 10 GK is related to the shape-phase transition. At this point, small differences between the rates are attributed to use of different effective interactions. As mentioned previously the "approx. mod." curve in Fig.~\ref{fig:compTotalRate}\hyperref[fig:compTotalRate]{(a),(b)} follows the original approximation~\cite{Langanke2003} bellow $T = 5$~GK for $\rho Y_e = 10^{11}$~g/cm$^{-3}$ and above $\rho Y_e = 10^{11}$~g/cm$^{-3}$ for $T=10$~GK, because of the validity range of the parametrization~\cite{Raduta2017}. The EC rate of the diamond region from both approximations are less sensitive to the temperature than temperature-dependent QRPA calculations. The latter give lower rates at temperatures $T\lesssim10$~GK and higher rates above $T\approx 30$~GK, for a density of $\rho Y_e = 10^{11}$~g/cm$^{-3}$, similar to the specific case of $^{86}$Kr.

Finally, all the new microscopic calculations of the EC rate of the diamond region agree within around one order of magnitude at $T=10$~GK and $\rho Y_e = 10^{11}$~g/cm$^{-3}$, conditions where the deleptonization is relatively important during the CCSN. Therefore, one can expect small variations in the dynamics of CCSN associated with the choice of microscopic calculations, this point is discussed in the next section.

\section{Core-collapse simulations}%
\label{sec:simulation}

In order to study the impact of the temperature dependent EC rates on the core-collapse dynamics we used the GR1D numerical simulation code. This code treats the collapse and the early stage of the post-bounce phase in spherical symmetry with general-relativistic hydrodynamics and neutrino-transport based on the NuLib neutrino-interaction library. Details about GR1D and NuLib can be found in~\cite{OConnor2010,OConnor2015,Sullivan2015}. The results presented in this section are obtained with a 15-solar-mass, solar-metallicity star progenitor (s15WW95,~\cite{Woosley1995}) and the tabulated nuclear statistical equilibrium equation of state SFHo~\cite{Steiner2013}. We compare five simulations with different EC rates for nuclei in the diamond region. Three simulations were performed with the new finite-temperature EC rates presented in this work sections~\ref{sec:nonrel_qrpa} and~\ref{sec:rel_ftqrpa} with and without including the first-forbidden transitions, as well as two simulations based on the EC rates parameterizations~\cite{Langanke2003,Raduta2017} used in the previous section.
\begin{figure}
\centering
\hspace*{-0.5cm}\includegraphics[width=.69\columnwidth]{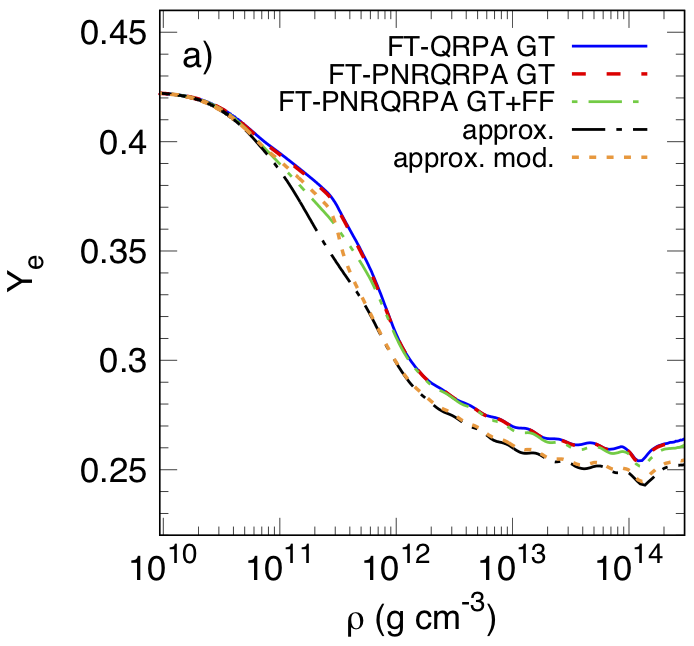}
\hspace*{-0.3cm}\includegraphics[width=.71\columnwidth]{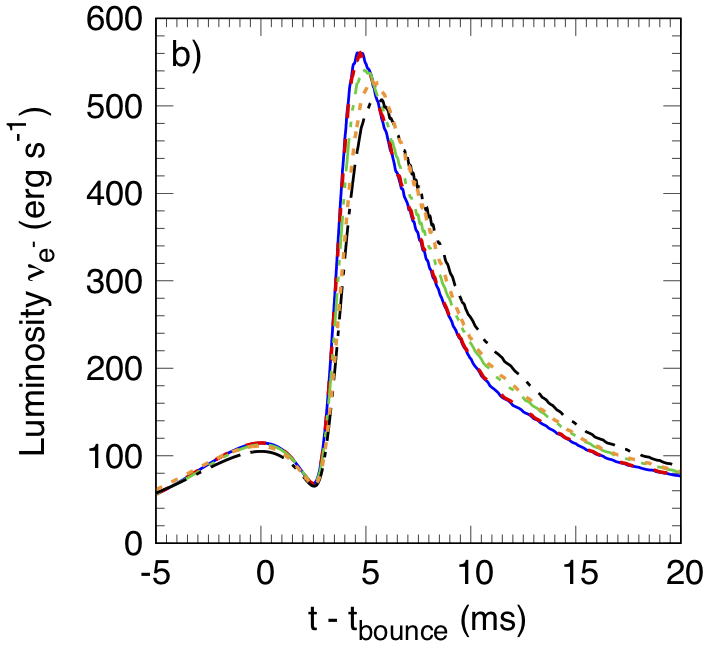}
\includegraphics[width=.7\columnwidth]{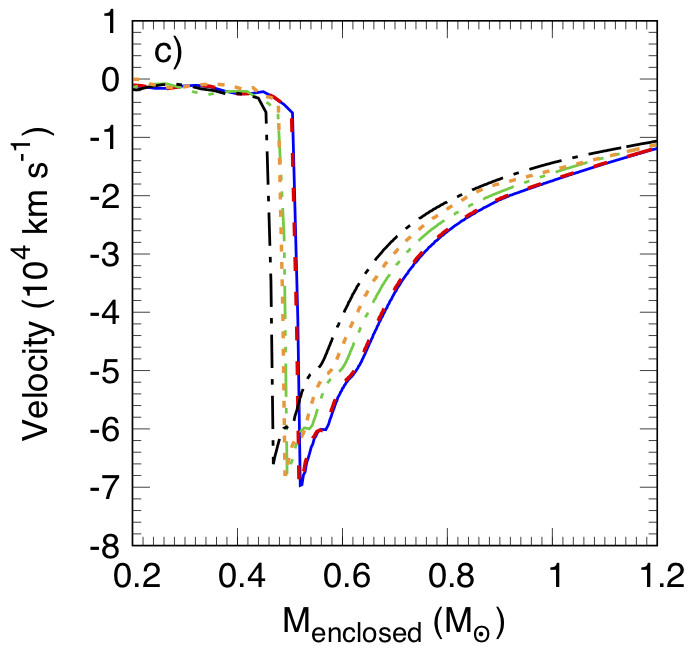}
\caption{\label{fig:yeLumVel} Core-collapse simulations results obtained from GR1D~\cite{OConnor2010,OConnor2015} and NULIB~\cite{Sullivan2015} codes, with s15WW95~\cite{Woosley1995} progenitor and SFHo~\cite{Steiner2013} equation of state. Different finite-temperature electron-capture calculations are compared, see labels in panel (a), where the three first are the finite-temperature microscopic calculations introduced in this paper, 
"approx." corresponds the parametrization~\cite{Langanke2003} and "approx. mod." stands for its modified version (third parametrization in~\cite{Raduta2017}). (a) the electron-fraction $Y_e$ as function of central baryon density $\rho$, (b) the electron neutrino luminosity as measured at a radius of 500 km as function of time after bounce and (c) the central velocity as function of the enclosed mass.}
\end{figure}

A comparison of the evolution of the electron fraction ($Y_e$) as function of the density of the inner core is shown in Fig.~\ref{fig:yeLumVel}\hyperref[fig:yeLumVel]{(a)}. We have shown previously in Sec.~\ref{sec:comparison} that the FT-QRPA and the FT-PNRQRPA calculations without first-forbidden transitions give similar EC rates, within a factor 10. No significant difference on the $Y_e$ evolution is observed when comparing these two sets. The effect of including the first-forbidden transitions in FT-PNRQRPA calculations is mostly notable for $8\times10^{10} \lesssim \rho \lesssim 8\times10^{11}$ g.cm$^{-3}$, conditions at which the most abundant nuclei are in the diamond region~\cite{Sullivan2015}. At $\rho Y_e = 3\times10^{11}$ g.cm$^{-3}$ and $T = 13.9$~GK, the $Y_e$ is reduced by $3 \%$ compared to the calculations with rates based on Gamow-Teller transitions only, while for these thermodynamics conditions the EC rates are about one order of magnitude higher when including first-forbidden transitions. The original approximation~\cite{Langanke2003} and its modified version~\cite{Raduta2017} lead to lower $Y_e$, because the EC rates of the nuclei populated during the deleptonization are overall higher than the EC rates from our finite-temperature microscopic calculations. 

In addition, the models with higher EC rates produce smaller electron-neutrino luminosity, Fig.~\ref{fig:yeLumVel}\hyperref[fig:yeLumVel]{(b)}, and lower homologous inner core mass, Fig.~\ref{fig:yeLumVel}\hyperref[fig:yeLumVel]{(c)}, as already discussed in~\cite{Sullivan2015,Pascal2020}. Including the first-forbidden transitions to the FT-PNRQRPA calculation reduces the amplitude of the main electron-neutrino luminosity burst by $3 \%$ and the mass of the homologous inner core by $4 \%$. Although this variation of homologous inner-core mass effects slightly the kinetic energy available for the shock wave, the description of the EC rates of nuclei in the diamond region is now better constrained by the new microscopic calculations presented in this work. 

The CCSN dynamics is not strongly dependent on the EC rate set used. The differences between the EC rate predictions have a relative small effect on the dynamics because at high EC rates the neutrino absorption increases and speeds up the onset of neutrino trapping, thus reducing the effective time of deleptonization from nuclei in the diamond region. Therefore, unlike a scenario where the EC rates are relatively low and sensitivities of the CCSN dynamics on variations in the rate are high, at high EC rates, such sensitivities are strongly reduced, as discussed earlier in~\cite{Sullivan2015,Titus2018}.

\section{Conclusion}
\label{sec:conclusion}
In this work, we have studied the temperature dependence of EC rates for nuclei near $N=50$ above $^{78}$Ni that play an important role in the collapse phase of CCSN~\cite{Sullivan2015,Pascal2020}. For this purpose, two sets of  newly-developed finite-temperature QRPA calculations of EC rates were performed at thermodynamic conditions relevant for core-collapse supernovae: one consists of a non-relativistic FT-QRPA based on an axially-deformed Skyrme functional (SkO' parametrization) and using the charge-changing finite amplitude method, the other consists of a relativistic FT-QRPA including nuclear pairing in the charge-exchange channel (FT-PNRQRPA) based on the relativistic nuclear energy density functional with momentum-dependent self-energies (D3C* parametrization). In the latter, both allowed (GT) and first-forbidden (FF) transitions have been included.

In addition, we have performed a large-scale shell-model calculations on $^{86}$Kr for better understanding the effects of finite-temperature on the EC rate of Pauli blocked nuclei at $N=50$. The main unblocking mechanism appears to be the thermal excitation of states for which the $g_{9/2}$ shell is occupied by at least one proton. The interplay between the nuclear structure effects, the electron-capture phase-space factor and the thermal population of initial states is complex and the  EC rate on $^{86}$Kr is dominated by GT transitions from a small group of excited states with negative-parity.

The comparison of the EC rates for $^{86}$Kr at $\rho Y_e = 10^{11}$~g.cm$^{-3}$, shows that the shell model predicts higher rates than finite-temperature QRPA models below $T\approx15$ GK, while the rates from the FT-QRPA models are higher above $T\approx15$ GK. The EC rates based on the shell model GT strengths are close to predictions from the parameterized approximations of~\cite{Langanke2003,Raduta2017} above $T\approx15$ GK. From comparisons of the rates on the neutron-rich nuclei of interest, and at thermodynamic conditions of CCSN, the two FT-QRPA GT-only calculations agree within a range of about an order of magnitude. The main discrepancies emerge around $^{79}$As ($Z=33$, $N=46$) and for the most neutron-rich nuclei. The agreement improves with increasing temperature as the rates depend less on the details of the nuclear structure. Finally, with the FT-PNRQRPA calculations we have shown that the contributions from the FF transitions are significant, especially at low temperature: the EC rates increase by about an order of magnitude for $T\lesssim10$~GK at $\rho Y_e = 10^{11}$~g.cm$^{-3}$. The results with FT-PNRQRPA including FF contributions are consistent with the results from~\cite{Dzhioev2020}.

Finally, the new finite-temperature electron-capture rates have been applied in 1D core-collapse simulations. Although the total EC rates for nuclei in the region of interest can vary  by an order of magnitude during the deleptonization phase, depending on the choice of the model used, the maximum electron-neutrino luminosity and the enclosed mass at core bounce are impacted by less than $5\%$. The new microscopic calculations presented in this work better constrain the EC rates and uncertainties can be better quantified. Therefore, the uncertainties introduced in core-collapse dynamical simulations due to uncertainties in EC rates are reduced and better understood. Nonetheless, the differences between the new finite-temperature EC rates could still have significant impacts on the scenarios of other astrophysical phenomena occurring at lower density, such as the thermal evolution of the neutron-star crust~\cite{Schatz2014,Chamel2021} and nucleosynthesis in thermonuclear supernovae~\cite{Iwamoto1999,Bravo2019}. It will be important to extend studies of the temperature dependence of EC rates to other regions of the chart of nuclei to investigate the impact on other astrophysical phenomena. Present theoretical models have proven to be instrumental in constraining the main observables of the CCSNe evolution. Theoretical calculations have now progressed to the point where models based on completely different assumptions and effective interactions (relativistic vs non relativistic FT-QRPA or shell-model) provide consistent description of EC rates, and produce reasonably small uncertainties in modeling the CCSNe. Therefore, we are now at the stage to perform large-scale calculations of the EC rates across the nuclide chart and establish a consistent table of EC rates available for the whole nuclear astrophysics community.

\begin{acknowledgments}
This work was supported by the US National Science Foundation under Grants PHY-1913554 (Windows on the Universe: Nuclear Astrophysics at the NSCL), PHY-1430152 (JINA Center for the Evolution of the Elements), PHY-1927130 (AccelNet-WOU: International Research Network for Nuclear Astrophysics [IReNA]) and PHY-2110365 (Nuclear Structure Theory and its Applications to Nuclear Properties, Astrophysics and Fundamental Physics). A.R. and N.P. acknowledge support by the QuantiXLie Centre of Excellence, a project co-financed by the Croatian Government and European Union through the European Regional Development Fund, the Competitiveness and Cohesion Operational Programme (KK.01.1.1.01.0004).
\end{acknowledgments}

% The \nocite command causes all entries in a bibliography to be printed out
% whether or not they are actually referenced in the text. This is appropriate
% for the sample file to show the different styles of references, but authors
% most likely will not want to use it.
%\nocite{*}

\clearpage
\newpage
\bibliographystyle{apsrev4-1}
\bibliography{bib}% Produces the bibliography via BibTeX.

\end{document}